\documentclass[reqno,11pt]{amsart}
\usepackage{multind}
\usepackage{hyperref}
\usepackage{graphicx}
\usepackage{amscd}
\usepackage{slashed}
\usepackage[mathscr]{eucal}
\numberwithin{equation}{section}
\allowdisplaybreaks[1]

\title[Bosonic Loop Diagrams in Classical Field Theory]{Bosonic Loop Diagrams
as Perturbative Solutions of the Classical Field Equations in $\phi^4$-Theory}

\author[F.\ Finster]{Felix Finster}
\address{Fakult\"at f\"ur Mathematik \\ Universit\"at Regensburg \\ D-93040 Regensburg \\ Germany}
\email{finster@ur.de}
\author[J.\ Tolksdorf]{J\"urgen Tolksdorf \\ \\ January 2012}
\address{Max Planck Institute for Mathematics in the Sciences, Leipzig, Germany}
\email{Juergen.Tolksdorf@mis.mpg.de}
\thanks{F.F.\ is supported in part by the Deutsche Forschungsgemeinschaft.
The research leading to these results has received funding from the
European Research Council under the European Union's Seventh Framework
Programme (FP7/2007-2013) / ERC grant agreement n$^\circ$~267087.}
\newtheorem{Def}{Definition}[section]

\newtheorem{Prp}[Def]{Proposition}
\newtheorem{Lemma}[Def]{Lemma}

\newcommand{\Thanks}{\vspace*{.5em} \noindent \thanks}
\newcommand{\beq}{\begin{equation}}
\newcommand{\eeq}{\end{equation}}
\newcommand{\Proof}{\begin{proof}}
\newcommand{\QED}{\end{proof} \noindent}

\newcommand{\la}{\langle}
\newcommand{\ra}{\rangle}
\newcommand{\bra}{\mathopen{\big<}}
\newcommand{\ket}{\mathclose{\big>}}

\newcommand{\C}{\mathbb{C}}
\newcommand{\R}{\mathbb{R}}
\newcommand{\1}{\mbox{\rm 1 \hspace{-1.05 em} 1}}
\newcommand{\Z}{\mathbb{Z}}

\renewcommand{\H}{\mathscr{H}}

\newcommand{\bep}{\begin{pmatrix}}
\newcommand{\enp}{\end{pmatrix}}

\renewcommand{\O}{\mathscr{O}}

\newcommand{\F}{{\mathscr{F}}}

\newcommand{\D}{{\mathfrak{D}}}

\DeclareMathOperator{\re}{Re}

\DeclareMathOperator{\supp}{supp}
\DeclareMathOperator{\Texp}{Texp}

\newcommand{\p}{\mathfrak{p}}

\begin{document}
\maketitle

\begin{abstract}
Solutions of the classical $\phi^4$-theory in Minkowski space-time are analyzed
in a perturbation expansion in the nonlinearity.
Using the language of Feynman diagrams, the solution of the Cauchy problem is expressed
in terms of tree diagrams which involve the retarded Green's function and have one outgoing leg.
In order to obtain general tree diagrams, we set up a ``classical measurement process''
in which a virtual observer of a scattering experiment modifies the field and detects suitable energy differences.
By adding a classical stochastic background field, we even obtain all loop diagrams.
The expansions are compared with the standard Feynman diagrams of the corresponding quantum
field theory.
\end{abstract}

\tableofcontents

\section{Introduction}
The dynamics of interacting quantum fields is most successfully described 
perturbatively in terms of Feynman diagrams.
One distinguishes between tree diagrams, which are finite, and loop diagrams, which typically diverge.
Performing a suitable ``classical limit''~$\hbar \rightarrow 0$, the loop diagrams drop out, and
only the tree diagrams remain. Therefore, it is a common opinion that tree diagrams are inherent
already in the classical theory, whereas loop diagrams describe the particular effects of quantized fields.
However, no hints can be found in the literature on how this statement could be made precise.
This was our motivation for approaching the question coming from the classical theory:
We consider a classical field, being a solution of a nonlinear hyperbolic partial differential
equation. Treating the nonlinearity perturbatively, we obtain an expansion which we express
in the language of Feynman diagrams. Then the question is, does the resulting expansion
contain all tree diagrams of quantum field theory? Are there differences in the tree expansions?
Do loop diagrams appear? Is it really impossible to obtain loop diagrams within the realm of
classical field theory? Our hope was that analyzing these questions would give a better understanding
of what the essence and physical significance of field quantization is.
In particular, we wanted to clarify how the nonlinear dynamics of classical fields
and the corresponding nonlinear scattering theory fit into the framework of
a linear dynamics of quantum fields on the Fock space.

Our results are unexpected in several ways. First, rewriting the solution of the Cauchy problem
in terms of Feynman diagrams, we only obtain diagrams with one outgoing leg,
in which all lines correspond to retarded Green's functions.
Several outgoing legs are obtained only by considering a scattering experiment
and setting up a ``classical measurement process'',
in which the wave is modified with an inhomogeneity at large times, and the difference of
the energies before and after this modification are measured. Furthermore, in order not to distinguish
a direction of time, we replace the perturbation expansion involving the retarded Green's function
by the so-called global perturbation expansion. This gives complete agreement with the
tree diagrams of quantum field theory, except that, instead of the Feynman propagator,
other, necessarily real-valued, Green's functions appear. In this sense, we
can confirm that tree diagrams are in fact inherent already in the classical theory.
However, this statement is true only in the special setting of the classical measurement process, and
only up to the differences in the choice of the propagators.

As another surprising result, we find that it is indeed possible to obtain loop diagrams in a
purely classical context. To this end, we consider the classical field in a stochastic background,
as described by a Gaussian ensemble of classical solutions. The resulting expansion in terms
of Feynman diagrams includes all loop diagrams of quantum field theory. Our expansion differs from that
of QFT in that the Feynman propagators are to be replaced by real-valued Green's functions and
fundamental solutions, with a specific combinatorics.
Moreover, there are differences in the combinatorial factors for higher order loop diagrams, as we explain
in a simple example.

Working with a stochastic field has some similarity with the approaches to explain quantum
effects by adding a stochastic term to the classical equations (see for example
Nelson's stochastic mechanics~\cite{nelson} or~\cite{pena-cetto, khrennikov}). However, in contrast to these approaches, we do not modify the classical equations.
We merely add a stochastic background field to the initial data. This additional field participates in the
classical interaction as described by the classical field equations. The nonlinearity of the mutual
interaction gives rise to the loop diagrams.
The physical picture is that the macroscopic field is superimposed by microscopic fluctuations,
which can be thought of as a classical analog of vacuum fluctuations.
These microscopic fluctuations can be observed only indirectly via their influence on the dynamics of the
macroscopic field.

We point out that our model of a classical field theory in a stochastic background field is clearly not
equivalent to a quantized field. In particular, our wave modes are classical, whereas in
quantum field theory they correspond to quantum mechanical oscillators.
But our point is that on the level of Feynman diagrams, these differences might not be visible.
In particular, the radiative corrections might not necessarily be a test for the quantum nature
of physical fields.

As is common in perturbative QFT, our treatment is formal in the sense that we do not
care about the divergences of diagrams and disregard all convergence issues of the expansions.
Thus all our expansions are to be understood as formal power series involving symbolic
expressions.
Moreover, in order to keep the setting as simple as possible, we always restrict attention
to the real massless $\phi^4$-theory in $(3+1)$-dimensional Minkowski space. But our methods and
results immediately extend to any other bosonic field theory in arbitrary dimension.

The paper is organized as follows. In Section~\ref{sec2} we provide the necessary background
on the Cauchy problem in classical field theory and on Feynman diagrams in QFT.
In Section~\ref{sec3} we consider expansions of the classical field in terms of Feynman diagrams.
Taking free fields as the starting point (Section~\ref{sec31}), we express the
solution of the Cauchy problem in terms of Feynman diagrams (Section~\ref{secnonlinear}).
We then set up the ``classical measurement process'' and derive an expansion in terms of
tree diagrams (Section~\ref{secmeasure}--\ref{seccscatter}).

Section~\ref{sec4} is devoted to the classical field theory in a stochastic background.
After introducing the free stochastic background field (Section~\ref{sec41}),
this field is included in the perturbation expansion to obtain loop diagrams (Section~\ref{sec42}).
The comparison to the loop diagrams in QFT is given in Section~\ref{sec43}.

Finally, in Section~\ref{sec5} we give an outlook on more realistic theories including fermions. We also
outline potential applications of our methods to constructive field theory.

\section{Preliminaries on $\phi^4$-Theory} \label{sec2}
We introduce classical $\phi^4$-theory in the Lagrangian formulation.
The Lagrangian~${\mathcal{L}}$ is given by
\[ {\mathcal{L}} = \frac{1}{2}\: (\partial_\mu \phi) (\partial^\mu \phi) - \frac{\lambda}{4!}\: \phi^4 \:, \]
where~$\phi$ is a real-valued scalar field.
Integrating the Lagrangian over Minkowski space-time gives the action~${\mathcal{S}}$,
\[ {\mathcal{S}} = \int {\mathcal{L}}(\phi, \partial \phi)\: d^4x \:. \]
We are working in units where~$c=1=\hbar$
and denote a chosen length scale by~$\ell$. Then the field has dimension of inverse length, whereas
the coupling constant~$\lambda$ and the classical action are dimensionless,
\[ [\phi] = \ell^{-1} \qquad \text{and} \qquad [\lambda] = \ell^0 = [{\mathcal{S}}] \:. \]
Considering critical points of the action, one obtains the Euler-Lagrange (EL) equations
\beq \label{cfe} \Box \phi = -\frac{\lambda}{6}\, \phi^3
\eeq
(where~$\Box = \partial_t^2 - \Delta_{\R^3}$ is the wave operator).

According to Noether's theorem, the symmetries of the Lagrangian correspond to
conserved quantities. In particular, the symmetry under time translations gives rise
to the conserved classical energy~$E$,
\beq \label{E0}
E(\phi) = \int_{t=T} 
\left( \frac{1}{2}\: \dot{\phi}^2 + \frac{1}{2}\: |\nabla \phi|^2 + \frac{\lambda}{4!}\:
\phi^4 \right) d^3x \:,
\eeq
having the units
\[ [E] = \ell^{-1} \:. \]

\subsection{Basics on the Classical Cauchy Problem and Scattering Theory} \label{sec21}
In the Cauchy problem, one seeks for solutions of the nonlinear hyperbolic
partial differential equation~\eqref{cfe}
for initial data~$(\phi, \dot{\phi})|_{t_0}$ 
given at some time~$t_0$.
It is a classical result that the Cauchy problem is well-posed, meaning that
there is a unique solution for short times in a suitable Sobolev space.
A general method for the proof is provided by the theory of symmetric hyperbolic
systems (see~\cite{john, taylor3}). The theory of symmetric hyperbolic systems also
reveals that the solutions of hyperbolic partial differential equations propagate with finite speed,
showing that the physical principle of causality is respected.
In general, the solution of the Cauchy problem will not exist for all times, as singularities may form. 
It has been a major topic of mathematical research to 
obtain global existence results and estimates of the solution for long times
(see for example the textbooks~\cite{john2, sogge, kichenassamy}).
Here we shall disregard all analytical questions by simply assuming that our solution
exists for all times.

In scattering theory, one is interested in the asymptotic behavior of the field in the limits~$t \rightarrow \pm \infty$.
Scattering theory has been developed mainly for linear equations
(for good references on linear scattering theory in the mathematical physics literature
see~\cite{reed+simon3, melrose, roach, derezinski+gerard, yafaev}).
In the standard setting, the dynamics is described by a linear evolution
equation of the form of the Schr\"odinger equation
\beq \label{lindyn}
i \partial_t \psi = H(t)\, \psi \:,
\eeq
where~$H(t)$ is a self-adjoint linear operator on a Hilbert space~$(\H(t), ( .,. ))$.
Then the corresponding time-evolution operator~$U(t_1, t_2)$ defined by
\[ U(t_1, t_2) \::\: \H(t_2) \rightarrow \H(t_1) \::\: \psi|_{t_2} \mapsto \psi|_{t_1} \]
is a unitary operator between Hilbert spaces. One wants to compare the
interacting dynamics~\eqref{lindyn} with a free dynamics of the form
\beq \label{lindynfree}
i \partial_t \psi = H_0(t)\, \psi\:,
\eeq
where~$H_0$ is a self-adjoint linear operator on a Hilbert space~$(\H_0, ( .,. ))$.
Usually, the operator~$H_0$ is so simple that the 
time evolution operator~$U_0(t_1, t_2)$ is explicitly known.
Then the {\em{wave operators}} are introduced by considering the late-time limits
\beq \label{Omegapm}
\Omega_\pm \psi := \lim_{t \rightarrow \pm\infty} U(t_0, t)\: U_0(t, t_0) \,\psi \::\: \H_0 \rightarrow \H
\equiv \H(t_0) \:,
\eeq
where for large times we identified the Hilbert spaces~$\H_0$ and~$\H(t)$.
Then the {\em{scattering operator}}~$S$ is defined by
\beq \label{Sclass}
S = (\Omega_-)^* \,\Omega_+ \::\: \H_0\rightarrow \H_0 \:;
\eeq
it maps the free outgoing field to the corresponding free incoming field.

As the $\phi^4$-theory is non-linear, the formalism of linear scattering does not immediately apply.
In order get a rigorous connection, one would need to prove that the wave dissipates for large times,
going over to a solution of the free wave equation. This requires decay estimates in the spirit
of~\cite{strauss, ginibre+velo}; see also~\cite[Chapter~9]{roach} and~\cite[Section~6]{strauss2}.
In this paper, we shall not enter dissipation and decay estimates, but instead we will restrict attention to a
systematic treatment of nonlinear classical scattering on the level of formal perturbation expansions.
We will enter this analysis in Section~\ref{sec3}. In order to get into the position to compare the
different settings, we now briefly summarize the basics on quantum scattering.

\subsection{Basics on Quantum Field Theory and Feynman Diagrams} \label{secpertqu}
We now review the basics on the corresponding $\phi^4$-quantum field theory.
For brevity, we keep the presentation on the level of formal calculations
(for attempts for making these calculations mathematically sound see for example~\cite{glimm+jaffe}).
We begin with the formalism of {\em{canonical quantization}}.
To quantize the free field, we first write the general solution of the classical equation~$\Box \phi = 0$
as a Fourier integral,
\beq \label{freeFourier}
\begin{split}
\phi_0(x) &= \int \frac{d^4 k}{(2 \pi)^4}\:\delta(k^2)\: \phi_0(k)\: e^{-i k x} \\
&= \frac{1}{(2 \pi)^4} \int \frac{d^3 k}{2 \omega} \left(
\phi_0(\omega, \vec{k})\, e^{-i \omega t + i \vec{k} \vec{x}}
+ \phi_0(-\omega, -\vec{k}) \, e^{i \omega t- i \vec{k} \vec{x}} \right) ,
\end{split}
\eeq
where we set~$\omega = |\vec{k}|$. The fact that our field is real means that the Fourier
coefficients satisfy the relation
\beq \label{freereal}
\overline{\phi_0(\omega, \vec{k})} = \phi_0(-\omega, -\vec{k})\:.
\eeq
For the quantization, the Fourier coefficients are replaced by operator-valued distributions
according to
\[ \phi_0(\omega, \vec{k}) \longrightarrow 2 \pi \,\sqrt{2 \omega} \,a(\vec{k})\:,\qquad
\phi_0(-\omega, -\vec{k}) \longrightarrow 2 \pi \,\sqrt{2 \omega} \, a^\dagger(\vec{k})\:, \]
where~$a(\vec{k})$ and~$a^\dagger(\vec{k})$ satisfy the commutation relations
\beq \label{ccr}
\big[ a(\vec{k}), a^\dagger(\vec{q}) \big] = (2 \pi)^3\, \delta^3(\vec{k}-\vec{q})\:,\qquad
\big[a(\vec{k}), a(\vec{q}) \big] = 0 = \big[ a^\dagger(\vec{k}), a^\dagger(\vec{q}) \big] \:.
\eeq
We denote the resulting quantized field by~$\hat{\phi}_0$,
\beq \label{phi0rep}
\hat{\phi}_0(x) = \int \frac{d^3 k}{(2 \pi)^3}\: \frac{1}{\sqrt{2 \omega}} \left(
a(\vec{k})\, e^{-i \omega t + i \vec{k} \vec{x}} + a^\dagger(\vec{k}) \, e^{i \omega t- i \vec{k} \vec{x}} \right) .
\eeq
As an immediate consequence of~\eqref{ccr} and~\eqref{phi0rep},
the quantized field has the following properties:
\begin{align}
\Box \hat{\phi}_0(x) &= 0 \label{freequant} \\
\big[ \hat{\phi}_0(t, \vec{x}), \hat{\phi}_0(t, \vec{y}) \big] &= 0 
= \big[ \partial_t \hat{\phi}_0(t, \vec{x}), \partial_t \hat{\phi}_0(t, \vec{y}) \big] \label{ccrphi1} \\
\big[ \partial_t \hat{\phi}_0(t, \vec{x}), \hat{\phi}_0(t, \vec{y}) \big] &= -i \delta^3(\vec{x}-\vec{y})\:. \label{ccrphi2}
\end{align}
Noting that~$\partial_t \hat{\phi}$ is the canonical momentum, equation~\eqref{ccrphi2} is the analog of
the commutation relation~$[p, q]=-i$ of quantum mechanics.
For general space-time points~$x$ and~$y$, a short calculation yields the
commutation relations
\beq \label{ccrxy}
\big[\hat{\phi}_0(x), \hat{\phi}_0(y) \big] = 
2 \pi K_0(x-y)\:,
\eeq
where~$K_0(x)$ is the distribution
\begin{align}
K_0(x) &= \int \frac{d^4k}{(2 \pi)^4}\: \delta(k^2)\: \epsilon(k^0)\: e^{-i k x} \label{K0def} \\
&= \frac{1}{(2 \pi)^4} \int \frac{d^3k}{2 \omega} \left(e^{-i \omega t + i \vec{k} \vec{x}}
- e^{i \omega t- i \vec{k} \vec{x}} \right) . \label{Krep}
\end{align}
We remark that the equal-time commutation relations~\eqref{ccrphi1} and~\eqref{ccrphi2}
immediately follow from~\eqref{ccrxy} and~\eqref{Krep} by 
differentiating with respect to time and taking the limit~$x^0-y^0 \rightarrow 0$.
The Fourier representation~\eqref{K0def} shows that~$K_0(x)$ is a distributional solution of the wave equation.
Moreover, this distribution is causal in the sense that it vanishes for spacelike~$x$.
This becomes apparent from the representation
\[ K_0 = \frac{1}{2 \pi i} \left( S^\vee - S^\wedge \right) , \]
where~$S^\vee$ and~$S^\wedge$ are the advanced and retarded Green's function defined by
\begin{align}
S^\wedge(x) &= \lim_{\varepsilon \searrow 0}
\int \frac{d^4 k}{(2 \pi)^4}\;
\frac{e^{-i k x}}{k^2 + i \varepsilon k^0} \label{Sret} \\
&= \left\{ \begin{array}{lcl} \displaystyle -i \int \frac{d^3 k}{(2 \pi)^3}\: \frac{1}{2 \omega}
\left( e^{-i \omega t + i \vec{k} \vec{x}} - e^{i \omega t + i \vec{k} \vec{x}} \right) &\quad& \text{if~$t>0$} \\[1em]
0 && \text{if~$t \leq 0$} \end{array} \right. \label{Sretrep} \\[.3em]
S^\vee(x) &= S^\wedge(-x)\:.
\end{align}

Usually, the operator algebra which is generated by~$a(\vec{k})$ and~$a^\dagger(\vec{k})$
is realized on a Fock space~$(\F, \la .|. \ra)$. This is a Hilbert space
with a distinguished normalized vector~$|0\ket$ (the ``vacuum state'')
having the following properties:
\begin{itemize}
\item[(i)] For all~$\vec{k}$, the operator~$a(\vec{k})$ annihilates the vacuum: $a(\vec{k}) |0\ra = 0$.
\item[(ii)] The vacuum is cyclic in the sense that by
iteratively applying the operators~$a^\dagger(\vec{k})$ to~$|0\ket$ one generates
a dense subspace of~$\F$.
\item[(iii)] For all~$\vec{k}$, the operator~$a^\dagger(\vec{k})$ is the formal adjoint of~$a(\vec{k})$
with respect to the scalar product~$\la .|. \ra$.
\end{itemize}
The properties~(i)--(iii) together with the commutation relations~\eqref{ccr}
determine the scalar product. For example, setting
\[ |\vec{k} \ra := a^\dagger(\vec{k}) |0 \ra
\qquad \text{and} \qquad \la \vec{k} |  := \big\la |\vec{k} \ra \big| \,.\, \big\ra \:, \]
we obtain
\begin{align*}
\la \vec{k} | \vec{q} \ra
&= \la 0 | a(\vec{k}) \,a^\dagger(\vec{q}) | 0 \ra
= \big\la 0 \big| [a(\vec{k}), a^\dagger(\vec{q})] \big| 0 \big\ra \\
&= (2 \pi)^3\, \delta^3(\vec{k}-\vec{q})\:  \la 0 | 0 \ra = (2 \pi)^3\, \delta^3(\vec{k}-\vec{q})\:.
\end{align*}
Thus, introducing for a Schwartz function~$f$ the vector
\beq \label{psidef}
| f \ra = \int \frac{d^3k}{(2 \pi)^3}\:\frac{1}{\sqrt{2 \omega}}\: f(\vec{k})\: | \vec{k} \ra \;\in\; \F\:,
\eeq
one obtains
\beq \label{psiscal}
\la f | g \ra = \frac{1}{(2 \pi)^3}\: \int \frac{d^3k}{2 \omega}\: \overline{f(\vec{k})} \: g(\vec{k})\:.
\eeq

In order to introduce the interaction, one wants to construct new field
operators~$\hat{\phi}$ on~$\F$ which solve the nonlinear classical field equation,
\beq \label{nonlinq}
\Box \hat{\phi}(x) = -\lambda\, \hat{\phi}(x)^3 \:.
\eeq
This can be accomplished by the ansatz
\beq \label{ans1}
\hat{\phi}(t, \vec{x}) = U(t)^\dagger \,\hat{\phi}_0(t, \vec{x}) \, U(t) \:,
\eeq
where~$U(t)$ is a formal solution of the equation
\beq \label{ans2}
i  \partial_t U(t) = -H_\text{int}(t)\: U(t) \qquad \text{with} \qquad
H_\text{int}(t) := \frac{\lambda}{4!} \int_{t=\text{const}} \hat{\phi}_0(t, \vec{x})^4\: d^3x\:.
\eeq
Indeed, applying the wave operator to~\eqref{ans1} and using~\eqref{ans2}
together with the commutation relations~\eqref{ccrphi1} and~\eqref{ccrphi2},
a straightforward calculation shows that~\eqref{nonlinq} is satisfied (for details see
for example~\cite[Section~3.5]{veltman}
in the similar context of a pion-field or~\cite[Chapter~4]{peskin+schroeder}).

Writing the interaction in the form~\eqref{ans1} and~\eqref{ans2}
is particularly useful for describing a {\em{quantum scattering process}}. To this end, we assume
that in the limits~$t \rightarrow \pm \infty$, the dynamics goes over to that of free fields.
Thus the asymptotic states are vectors of the free Fock space~$(\F, \la .|. \ra)$.
We denote the free field operators in the asymptotic regions
by~$\hat{\phi}_\text{in}$ and~$\hat{\phi}_\text{out}$
(note that they are solutions of the free field equation~\eqref{freequant}).
According to common notation, the Fock space~$(\F, \la .|. \ra)$
with the basis generated by~$a^\dagger_\text{in}$ and~$a^\dagger_\text{out}$
is denoted by~$(\F_\text{in}, \la .|. \ra_\text{in})$
and~$(\F_\text{out}, \la .|. \ra_\text{out})$, respectively. By construction, these
two Fock spaces are identical, but they come with different orthonormal bases.
In order to describe the transformation of one basis to the other,
one considers the solution of the above equations~\eqref{ans1} and~\eqref{ans2}
with~$\hat{\phi}_0 = \hat{\phi}_\text{in}$ and
\beq \label{Uboundary}
\lim_{t \rightarrow -\infty} U(t) = \1\:.
\eeq
Then~$U(t)$ can be given in terms of a time-ordered exponential,
\begin{align*}
U(t) &= \Texp \left( i \int_{-\infty}^t H_\text{int}(\tau)\: d\tau \right)
= \Texp \left( \frac{i \lambda}{4!}  \int_{ \{ y^0 < t \} }   \hat{\phi}_\text{in}(y)^4 \:d^4y \right) \\
&= \sum_{n=0}^\infty \frac{1}{n!} \left(\frac{i \lambda}{4!} \right)^n 
\int_{ \{ y_1^0 < t\} } d^4y_1 \cdots \int_{ \{ y_n^0 < t\} } d^4 y_n \: T \left( \hat{\phi}_\text{in}(y_1)^4\:\cdots\:
\hat{\phi}_\text{in}(y_n)^4 \right) .
\end{align*}
The scattering operator~$S$ is defined by
\begin{align}
S &:= \lim_{t \rightarrow \infty} U(t) \label{Squant} \\
&\,= \sum_{n=0}^\infty \frac{1}{n!} \left(\frac{i \lambda}{4!} \right)^n 
\int d^4y_1 \cdots \int d^4 y_n \: T \left( \hat{\phi}_\text{in}(y_1)^4\:\cdots\:
\hat{\phi}_\text{in}(y_n)^4 \right) .
\end{align}
Then, according to~\eqref{ans1},
\[ \hat{\phi}_\text{out} = S^\dagger \,\hat{\phi}_\text{in}\, S\:. \]
Moreover, noting that~$H_\text{int}$ is a symmetric operator on the Fock space,
one sees that~$S$ is a unitary operator on~$\F$,
\begin{align*}
S \::\: \F &\rightarrow \F \text{ unitary} \\ |\text{out} \ra &\mapsto |\text{in} \ra \:.
\end{align*}
We remark that this construction is very similar to the procedure in
classical linear scattering theory. While the scattering operator in the linear classical
setting~\eqref{Sclass} is a linear operator on the Hilbert space~$\H_0$ of classical fields,
the quantum scattering operator~\eqref{Squant} is a linear operator on the Fock space~$\F$.
Since here we are working in the interaction picture, the free dynamics~$U_0$
is the identity. Then the relations~\eqref{Omegapm} and~\eqref{Sclass} reduce
to~$S = U(-\infty, \infty)$. This is consistent with~\eqref{Squant} if in view
of~\eqref{Uboundary} we identify~$U(t)$ with~$U(-\infty, t)$.

The scattering amplitude of a scattering process can be expressed as
a matrix element of the scattering operator with respect to the in-basis,
\begin{align}
\la \beta_\text{out} \,|\, \alpha_\text{in} \ra &= \la \beta_\text{in} \,| S |\, \alpha_\text{in} \ra \label{matrix} \\
&= \sum_{n=0}^\infty \frac{1}{n!} \left(-\frac{i \lambda}{4!} \right)^n 
\int d^4y_1 \cdots \int d^4 y_n \: \Big\la \beta_\text{in} \, \Big|\, T \left( \hat{\phi}_\text{in}(y_1)^4\:\cdots\:
\hat{\phi}_\text{in}(y_n)^4 \right) \Big|\, \alpha_\text{in} \Big\ra \notag \\
&= \la \beta_\text{in} \,|\, \alpha_\text{in} \ra
-\frac{i \lambda}{4!} \int d^4y \:\big\la \beta_\text{in} \, \big|\, \hat{\phi}_\text{in}(y)^4
\big|\, \alpha_\text{in} \big\ra + \O(\lambda^2) . \notag
\end{align}
Generating the in-basis by iteratively applying~$\hat{\phi}_\text{in}$ to the vacuum (LSZ reduction
formalism; see for example~\cite{peskin+schroeder}), all expectation values can be expressed in terms
of the $n$-point functions defined by
\[ G_0(x_1, \ldots, x_n)  = \bra 0 \big| T \big( \hat{\phi}_\text{in}(x_1) \cdots \hat{\phi}_\text{in}(x_n) \big) \big|
0 \ket \:. \]
Using the commutation relations, the $n$-point functions can be written as
sums of products of the two-point functions (Wick's theorem).
A short computation using~\eqref{ccr} and~\eqref{phi0rep} yields for the
two-point function
\beq \label{GFeyn}
G_0(x,y) = i \triangle_F(x-y) \:,
\eeq
where~$\triangle_F$ is the Feynman propagator,
\beq \label{Feynprop} \begin{split}
\triangle_{F}(x) &:= \lim_{\varepsilon \searrow 0}
\int \!\frac{d^4k}{(2\pi)^4}\frac{e^{-ik x}}{k^2 + i \varepsilon} \\
&= -i \int \frac{d^3k}{(2 \pi)^3} \:\frac{1}{2 \omega} \left( \Theta(t)\: 
e^{-i \omega t + i \vec{k} \vec{x}} + \Theta(-t)\: e^{i \omega t - i \vec{k} \vec{x}} \right) .
\end{split}
\eeq
The Feynman propagator satisfies the defining equation of a Green's function
\[ -\Box \triangle_{F}(x) = \delta^4(x)\:. \]
However, in contrast to the advanced and retarded Green's function, it is
complex-valued, and it is non-zero in space-like directions.

For the systematic treatment of perturbation theory, it is most convenient to work with a generating functional
in the {\em{path integral formalism}}. We thus introduce the generating functional
\[ Z_\lambda[j] = \int e^{i ( \mathcal{S}[\phi] + j \cdot \phi )}\: {\mathscr{D}} \phi
\qquad \text{where} \qquad
j \cdot \phi := \int j(x)\, \phi(x)\:d^4x \:. \]
The generating functional is related to the vacuum expectation value of
canonical quantization by (see for example~\cite{weinberg} and~\cite{itzykson/zuber})
\[ Z_\lambda[j] = \Big\la 0 \:\Big| \Texp \left( - \frac{i \lambda}{4!} \int \hat{\phi}_\text{in}^4(x)\: d^4x
+i \int j(x)\: \hat{\phi}_\text{in}(x)\: d^4x \right) \Big|\: 0 \Big\ra\:. \]
In order to compute~$Z_\lambda[j]$, we rewrite the generating functional as
\beq \label{Zlam}
Z_\lambda[j] = \exp \left( - \frac{i \lambda}{4!} \int d^4x\: \Big(-i \frac{\delta}{\delta j(x)} \Big)^4 \right) Z_0[j]\:,
\eeq
where~$Z_0[j]$ is the generating functional of the free quantum field theory:
\begin{align}
Z_0[j] &:= \lim_{\varepsilon \searrow 0}
\int {\mathscr{D}}\phi
\: \exp \left( i\int \Big( -\frac{1}{2}\: \phi(x)\: ((\Box - i \varepsilon)\phi(x)) + j(x)\, \phi(x) \Big) d^4x \right) \notag \\
&= \lim_{\varepsilon \searrow 0}
\int {\mathscr{D}}\mu_\varepsilon(\phi)\:e^{i ({\mathcal{S}}_0[\phi] + j \cdot \phi)} \:. \label{Z0path}
\end{align}
Here~${\mathcal{S}}_0$ is the free action,
\[ {\mathcal{S}}_0[\phi] =  \frac{1}{2} \int (\partial_\mu \phi)(x)\: (\partial^\mu \phi)(x)\: d^4x \:, \]
and~${\mathscr{D}}\mu_\varepsilon(\phi)$ is the Gaussian measure
\[ {\mathscr{D}} \mu_\varepsilon(\phi) = e^{- \varepsilon \int \phi(x)^2\, d^4x}\: {\mathscr{D}} \phi\:. \]
This path integral can be computed explicitly to obtain
\beq \label{Z0}
Z_0[j] = \exp \left( -\frac{i}{2} \iint j(x)\: \triangle_F(x-y)\: j(y)\: d^4x\, d^4y \right)
\eeq
(we use the convention~$Z_0[0]=1$).
In particular,
\[ G_0(x,y) = -\frac{\delta}{\delta j(x)}\frac{\delta}{\delta j(y)}\:
Z_0[j] \Big|_{j=0} = i \triangle_F(x-y)\:, \]
in agreement with~\eqref{GFeyn}. Wick's theorem can be expressed as
\beq \begin{split}
G_0(x_1, \ldots, x_n) &= (-i)^n\: \frac{\delta}{\delta j(x_1)} \cdots \frac{\delta}{\delta j(x_n)}\:
Z_0[j] \big|_{j=0} \\
&= (-i)^n\: \frac{\delta}{\delta j(x_1)} \cdots \frac{\delta}{\delta j(x_n)}\:
e^{-\frac{i}{2} \iint j(x)\: \triangle_F(x-y)\: j(y)\: d^4x\, d^4y} \big|_{j=0}\:, 
\end{split} \label{wick}
\eeq
showing that the $n$-point functions can indeed be expressed as sums of products of the Feynman
propagator.

\begin{figure}
\begin{picture}(0,0)%
\includegraphics{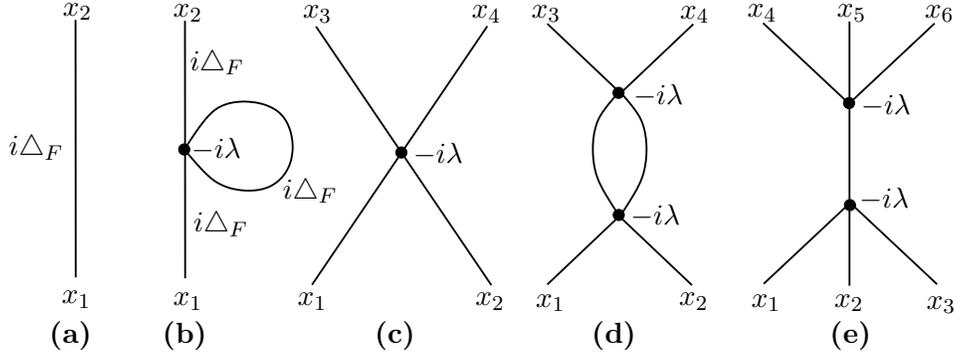}%
\end{picture}%
\setlength{\unitlength}{1533sp}%
\begingroup\makeatletter\ifx\SetFigFont\undefined%
\gdef\SetFigFont#1#2#3#4#5{%
  \reset@font\fontsize{#1}{#2pt}%
  \fontfamily{#3}\fontseries{#4}\fontshape{#5}%
  \selectfont}%
\fi\endgroup%
\begin{picture}(15054,5797)(-5024,-5147)
\put(-2054,-2011){\makebox(0,0)[lb]{\smash{{\SetFigFont{11}{13.2}{\rmdefault}{\mddefault}{\updefault}$-i \lambda$}}}}
\put(-2024,-3211){\makebox(0,0)[lb]{\smash{{\SetFigFont{11}{13.2}{\rmdefault}{\mddefault}{\updefault}$i \triangle_F$}}}}
\put(-2084,-586){\makebox(0,0)[lb]{\smash{{\SetFigFont{11}{13.2}{\rmdefault}{\mddefault}{\updefault}$i \triangle_F$}}}}
\put(-584,-2671){\makebox(0,0)[lb]{\smash{{\SetFigFont{11}{13.2}{\rmdefault}{\mddefault}{\updefault}$i \triangle_F$}}}}
\put(-4139,299){\makebox(0,0)[lb]{\smash{{\SetFigFont{11}{13.2}{\rmdefault}{\mddefault}{\updefault}$x_2$}}}}
\put(-2354,269){\makebox(0,0)[lb]{\smash{{\SetFigFont{11}{13.2}{\rmdefault}{\mddefault}{\updefault}$x_2$}}}}
\put(-2354,-4396){\makebox(0,0)[lb]{\smash{{\SetFigFont{11}{13.2}{\rmdefault}{\mddefault}{\updefault}$x_1$}}}}
\put(-4154,-4366){\makebox(0,0)[lb]{\smash{{\SetFigFont{11}{13.2}{\rmdefault}{\mddefault}{\updefault}$x_1$}}}}
\put(-5009,-1966){\makebox(0,0)[lb]{\smash{{\SetFigFont{11}{13.2}{\rmdefault}{\mddefault}{\updefault}$i \triangle_F$}}}}
\put(-339,-4401){\makebox(0,0)[lb]{\smash{{\SetFigFont{11}{13.2}{\rmdefault}{\mddefault}{\updefault}$x_1$}}}}
\put(2536,-4391){\makebox(0,0)[lb]{\smash{{\SetFigFont{11}{13.2}{\rmdefault}{\mddefault}{\updefault}$x_2$}}}}
\put(-4354,-5006){\makebox(0,0)[lb]{\smash{{\SetFigFont{11}{13.2}{\rmdefault}{\mddefault}{\updefault}{\bf{(a)}}}}}}
\put(1531,-2091){\makebox(0,0)[lb]{\smash{{\SetFigFont{11}{13.2}{\rmdefault}{\mddefault}{\updefault}$-i \lambda$}}}}
\put(5091,-1136){\makebox(0,0)[lb]{\smash{{\SetFigFont{11}{13.2}{\rmdefault}{\mddefault}{\updefault}$-i \lambda$}}}}
\put(5846,244){\makebox(0,0)[lb]{\smash{{\SetFigFont{11}{13.2}{\rmdefault}{\mddefault}{\updefault}$x_4$}}}}
\put(3511,-4426){\makebox(0,0)[lb]{\smash{{\SetFigFont{11}{13.2}{\rmdefault}{\mddefault}{\updefault}$x_1$}}}}
\put(5831,-4401){\makebox(0,0)[lb]{\smash{{\SetFigFont{11}{13.2}{\rmdefault}{\mddefault}{\updefault}$x_2$}}}}
\put(3441,259){\makebox(0,0)[lb]{\smash{{\SetFigFont{11}{13.2}{\rmdefault}{\mddefault}{\updefault}$x_3$}}}}
\put(5046,-3086){\makebox(0,0)[lb]{\smash{{\SetFigFont{11}{13.2}{\rmdefault}{\mddefault}{\updefault}$-i \lambda$}}}}
\put(8774,-1309){\makebox(0,0)[lb]{\smash{{\SetFigFont{11}{13.2}{\rmdefault}{\mddefault}{\updefault}$-i \lambda$}}}}
\put(8744,-2816){\makebox(0,0)[lb]{\smash{{\SetFigFont{11}{13.2}{\rmdefault}{\mddefault}{\updefault}$-i \lambda$}}}}
\put(8336,-4423){\makebox(0,0)[lb]{\smash{{\SetFigFont{11}{13.2}{\rmdefault}{\mddefault}{\updefault}$x_2$}}}}
\put(6998,-4426){\makebox(0,0)[lb]{\smash{{\SetFigFont{11}{13.2}{\rmdefault}{\mddefault}{\updefault}$x_1$}}}}
\put(9831,-4451){\makebox(0,0)[lb]{\smash{{\SetFigFont{11}{13.2}{\rmdefault}{\mddefault}{\updefault}$x_3$}}}}
\put(8381,266){\makebox(0,0)[lb]{\smash{{\SetFigFont{11}{13.2}{\rmdefault}{\mddefault}{\updefault}$x_5$}}}}
\put(9806,259){\makebox(0,0)[lb]{\smash{{\SetFigFont{11}{13.2}{\rmdefault}{\mddefault}{\updefault}$x_6$}}}}
\put(6933,274){\makebox(0,0)[lb]{\smash{{\SetFigFont{11}{13.2}{\rmdefault}{\mddefault}{\updefault}$x_4$}}}}
\put(-259,256){\makebox(0,0)[lb]{\smash{{\SetFigFont{11}{13.2}{\rmdefault}{\mddefault}{\updefault}$x_3$}}}}
\put(2476,254){\makebox(0,0)[lb]{\smash{{\SetFigFont{11}{13.2}{\rmdefault}{\mddefault}{\updefault}$x_4$}}}}
\put(901,-5006){\makebox(0,0)[lb]{\smash{{\SetFigFont{11}{13.2}{\rmdefault}{\mddefault}{\updefault}{\bf{(c)}}}}}}
\put(-2527,-5006){\makebox(0,0)[lb]{\smash{{\SetFigFont{11}{13.2}{\rmdefault}{\mddefault}{\updefault}{\bf{(b)}}}}}}
\put(4411,-5006){\makebox(0,0)[lb]{\smash{{\SetFigFont{11}{13.2}{\rmdefault}{\mddefault}{\updefault}{\bf{(d)}}}}}}
\put(8266,-5006){\makebox(0,0)[lb]{\smash{{\SetFigFont{11}{13.2}{\rmdefault}{\mddefault}{\updefault}{\bf{(e)}}}}}}
\end{picture}%
\caption{Feynman diagrams for the interacting $n$-point functions.}
\label{figloop1}
\end{figure}

The scattering amplitudes are most conveniently expressed in terms of
the {\em{interacting $n$-point functions}}~$G_\lambda$, which are
obtained by taking functional derivatives of the generating functional with interaction,
\beq \label{Glambda}
G_\lambda(x_1, \ldots, x_n) := 
 (-i)^n\: \frac{\delta}{\delta j(x_1)} \cdots \frac{\delta}{\delta j(x_n)}\,
Z_\lambda[j] \big|_{j=0}\:,
\eeq
Their perturbation expansion is again performed
by expanding with respect to~$\lambda$, \eqref{Zlam}, using~\eqref{Z0} and
applying the Wick rules. {\em{Feynman diagrams}}
are the pictorial representation of the perturbation expansion for the interacting $n$-point functions. They involve tree and loop diagrams; see Figure~\ref{figloop1} for a few examples.
A diagram is called {\em{connected}} if all outer lines are connected to each other.
In the functional calculus, the connected diagrams are represented by the
generating functional~$W_\lambda(j)=\log Z_\lambda(j)$. With this in mind,
in what follows we shall always restrict attention to connected diagrams.
For more details and further reading, we refer for example to~\cite{itzykson/zuber, weinberg, zee, veltman}.

For the later comparison with the classical theory, we now give the combinatorics of the diagrams.
We use the short notation
\beq \label{notfirst}
\triangle_F(j,j) \equiv \iint \triangle_F(x-y)\: j(x)\, j(y)\: d^4x\, d^4y\:.
\eeq
\begin{Lemma} \label{lemmacombi}
Suppose that we consider a contribution to~$G_\lambda(x_1, \ldots, x_n)$
of order~$k$ in perturbation theory. Then~$G_\lambda$ vanishes unless~$n$ is even.
Then the contribution can be represented by a Feynman diagram involving
\beq \label{Nrel}
N = \frac{n}{2} + 2 k
\eeq
lines. Analytically, the contribution can be written as
\beq \label{contri}
(-i \lambda)^k \: i^N \int d^4y_1\: \cdots \int d^4y_k\;
\prod_{\ell=1}^N \:\triangle_F(\ldots)\:.
\eeq
For connected tree diagrams, we have the additional relation
\beq \label{treerel}
k = \frac{n}{2} -  1\:.
\eeq
\end{Lemma}
\Proof According to~\eqref{Glambda}, \eqref{Zlam} and~\eqref{Z0},
\begin{align*}
&G_\lambda(x_1, \ldots, x_n) = 
 (-i)^n\: \frac{\delta}{\delta j(x_1)} \cdots \frac{\delta}{\delta j(x_n)}\, \\
 & \times \!\frac{1}{k!} \left( - \frac{i \lambda}{4!} \right)^k \int d^4y_1 \Big(-i \frac{\delta}{\delta j(y_1)} \Big)^4 
\!\!\cdots \int d^4y_k \Big(-i \frac{\delta}{\delta j(y_k)} \Big)^4
\frac{1}{N!} \left( -\frac{i}{2} \: \triangle_F(j,j) \right)^N_{\big| j=0} .
\end{align*}
For this contribution to be non-zero, we must have as many derivatives~$\delta/\delta j$
as factors~$j$. This gives~\eqref{Nrel}. Carrying out the derivatives, all the factorials
and factors~$1/2$ are compensated by combinatorial factors. Collecting the factors of~$-i$
and using~\eqref{Nrel} gives~\eqref{contri}.

To prove~\eqref{treerel} we note that in the case~$k=0$, only the two-point function contributes.
Obviously, each vertex increases~$n$ by two.
\QED

We finally recall how to take the ``classical limit'' in which only the tree diagrams remain.
As we want to take the limit~$\hbar \rightarrow 0$, we clearly need to work in
more general units where~$\hbar \neq 1$ (but still~$c=1$). Denoting the length scale
again by~$\ell$ and the mass scale by~$m$, all objects have dimensions of~$\ell$ and~$m$;
more precisely,
\beq \label{genunit}
[S] = [\hbar] = m \,\ell \:,\qquad [\phi] = \sqrt{\frac{m}{\ell}} \:,\qquad [\lambda] = \frac{1}{m \,\ell} \:,\qquad
[E] = m \:.
\eeq
Inserting a factor of~$1/\hbar$ into the exponent in~\eqref{Z0path}
and computing the path integral, one sees that~\eqref{Z0} is to be modified to
\beq \label{Z0hbar}
Z_0[j] = \exp \left( -\frac{i}{2 \hbar} \iint j(x)\: \triangle_F(x-y)\: j(y)\: d^4x\, d^4y \right) .
\eeq
Defining the interacting $n$-point functions by
\beq \label{Glhbar}
G_\lambda(x_1, \ldots, x_n) := 
\int \phi(x_1) \cdots \phi(x_n)\: e^{\frac{i}{\hbar}\,S}\: \D\phi \:,
\eeq
a short calculation using~\eqref{Z0hbar} shows that~\eqref{contri} is to be replaced by
\[ \hbar^{n+3k-N} (-i \lambda)^k \: i^N \int d^4y_1\: \cdots \int d^4y_k\;
\prod_{\ell=1}^N \:\triangle_F(\ldots) \]
(note that every factor~$\phi$ gives
the functional derivative~$\hbar\, \delta/\delta j$ and that expanding the exponential
in~\eqref{Glhbar} gives powers of~$\lambda/\hbar$). Using~\eqref{Nrel}, we conclude that
\beq \label{Gscal}
G_\lambda(x_1, \ldots, x_n) \sim \hbar^{k+\frac{n}{2}}\:.
\eeq
For connected tree diagrams, we conclude in view of~\eqref{treerel} that~$G_\lambda(x_1, \ldots, x_n)
\sim \hbar^{n-1}$. In order to understand this scaling, one must keep in mind that for describing a scattering
process, one needs to replace the outer lines by the incoming or outgoing fields.
Truncating the diagrams gives a factor~$\hbar^{-n}$ (because in view of~\eqref{Gscal} every
outer line~$G_0(x,y)$ carries a factor~$\hbar$). Hence the truncated diagrams
all scale~$\sim \hbar^{-1}$. After rescaling the diagrams by a common factor~$\hbar$, in the
connected tree diagrams all factors of~$\hbar$ drop out.
But~\eqref{Gscal} also shows that every loop gives rise to an additional factor~$\hbar$.
Thus taking the limit~$\hbar \rightarrow 0$, precisely the tree diagrams remain.

\section{A Nonlinear Classical Scattering Process} \label{sec3}
\begin{figure}
\begin{picture}(0,0)%
\includegraphics{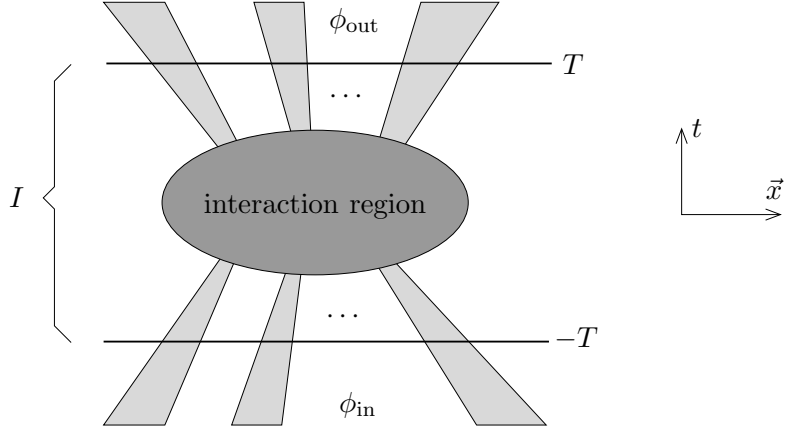}%
\end{picture}%
\setlength{\unitlength}{1533sp}%
\begingroup\makeatletter\ifx\SetFigFont\undefined%
\gdef\SetFigFont#1#2#3#4#5{%
  \reset@font\fontsize{#1}{#2pt}%
  \fontfamily{#3}\fontseries{#4}\fontshape{#5}%
  \selectfont}%
\fi\endgroup%
\begin{picture}(12532,6864)(-1994,-6013)
\put(1171,-2581){\makebox(0,0)[lb]{\smash{{\SetFigFont{11}{13.2}{\rmdefault}{\mddefault}{\updefault}interaction region}}}}
\put(6976,-331){\makebox(0,0)[lb]{\smash{{\SetFigFont{11}{13.2}{\rmdefault}{\mddefault}{\updefault}$T$}}}}
\put(6841,-4741){\makebox(0,0)[lb]{\smash{{\SetFigFont{11}{13.2}{\rmdefault}{\mddefault}{\updefault}$-T$}}}}
\put(-1979,-2476){\makebox(0,0)[lb]{\smash{{\SetFigFont{11}{13.2}{\rmdefault}{\mddefault}{\updefault}$I$}}}}
\put(9061,-1351){\makebox(0,0)[lb]{\smash{{\SetFigFont{11}{13.2}{\rmdefault}{\mddefault}{\updefault}$t$}}}}
\put(10261,-2341){\makebox(0,0)[lb]{\smash{{\SetFigFont{11}{13.2}{\rmdefault}{\mddefault}{\updefault}$\vec{x}$}}}}
\put(3196,-796){\makebox(0,0)[lb]{\smash{{\SetFigFont{11}{13.2}{\rmdefault}{\mddefault}{\updefault}$\cdots$}}}}
\put(3151,-4351){\makebox(0,0)[lb]{\smash{{\SetFigFont{11}{13.2}{\rmdefault}{\mddefault}{\updefault}$\cdots$}}}}
\put(3361,-5791){\makebox(0,0)[lb]{\smash{{\SetFigFont{11}{13.2}{\rmdefault}{\mddefault}{\updefault}$\phi_\text{in}$}}}}
\put(3196,419){\makebox(0,0)[lb]{\smash{{\SetFigFont{11}{13.2}{\rmdefault}{\mddefault}{\updefault}$\phi_\text{out}$}}}}
\end{picture}%
\caption{A nonlinear classical scattering process.}
\label{figscatter}
\end{figure}

In this section we shall analyze a scattering process for a classical field~$\phi$
being a solution of the classical field equation~\eqref{cfe}.
The physical situation which we have in mind is shown in Figure~\ref{figscatter}.
We have an incoming wave~$\phi_\text{in}$, which may be composed of several wave
packets. We assume for convenience
that for large negative times, the wave packets become more spread out in space, implying
that their amplitude becomes smaller. This has the advantage that in the limit~$t \rightarrow -\infty$,
$\phi$ can be treated as a free field (making it unnecessary to ``switch off'' the interaction
adiabatically). Similarly, for large positive times, $\phi$ should evolve into a free field~$\phi_\text{out}$.
We assume that the interaction
takes place inside the strip~$I=[-T, T] \times \R^3$ for a given large time~$T$ (in the end, we will
take the limit~$T \rightarrow \infty$). Inside the interaction region, $\phi$ satisfies the
nonlinear field equation~\eqref{cfe}, whereas outside the interaction region, $\phi$ is a
solution of the linear wave equation.

In the following subsections we first summarize how the Hilbert space of classical fields can be 
regarded as a subspace of the quantum Fock space. Afterwards, we discuss the perturbation expansion
of the classical (non-linear) Cauchy problem. The corresponding pictorial representation of this expansion
significantly differs from Feynman diagrams obtained in perturbative quantum field theory.  To
overcome this difference we motivate and introduce the so-called ``nonlinear classical measurement process".
This will make it possible to introduce the {\it classical scattering operator} and the {\it classical $n$-point functions}. The pictorial description of the perturbative expansion of these classical $n-$point
functions is then compared with the tree-diagrams of the corresponding quantum field theory. 

\subsection{The Free Classical Field} \label{sec31}
In preparation, we need to describe free fields, being solutions of the linear wave equation
\beq \label{freewave}
\Box \phi_0 = 0 \:.
\eeq
We assume for simplicity that~$\phi_0$ is smooth and spatially compact (note that, due to
finite propagation speed, the last property is preserved under the time evolution),
\[ \phi_0 \in C^\infty(\R, C^\infty_0(\R^3)) \:. \]
For free fields, the coupling constant~$\lambda$ vanishes, and thus the classical
energy~\eqref{E0} is quadratic. Thus by polarization we can introduce a scalar product,
\beq \label{E0free}
( \phi_0 , \psi_0 ) := \frac{1}{2} \int_{t=\text{const}} 
\left( \dot{\phi}_0 \dot{\psi}_0 + \vec{\nabla} \phi_0 \cdot \vec{\nabla} \psi_0 \right) d^3x \:.
\eeq
Due to energy conservation, this scalar product is independent of~$t$.
It makes the space of free solutions to a Hilbert space~$(\H_0, ( .,. ))$.

It will be useful to represent the function~$\phi_0$ in various ways.
First, in view of the unique Cauchy development, we can describe~$\phi_0$ by
its initial values at any given time~$t$,
\beq \label{intval}
\Phi(t) := (\phi_0, \partial_t \phi_0)|_{t} \in C^\infty_0(\R^3) \times C^\infty_0(\R^3) \:.
\eeq
Next, we can represent~$\phi_0$ similar to~\eqref{freeFourier} and~\eqref{freereal}
as a Fourier integral supported on the upper and lower mass cone,
\beq \label{Fourierrep}
\phi_0(x) = \frac{1}{(2 \pi)^4} \int \frac{d^3 k}{2 \omega} \left(
\phi_0(\vec{k})\, e^{-i \omega t + i \vec{k} \vec{x}}
+ \overline{\phi_0(\vec{k})} \, e^{i \omega t- i \vec{k} \vec{x}} \right)
\eeq
where again~$\omega = |\vec{k}|$.
Then the energy scalar product~\eqref{E0free} becomes
\beq
( \phi_0 , \phi_0 ) = \frac{1}{(2 \pi)^4}
\int \frac{d^3k}{2 \omega}\: \frac{\omega}{2 \pi}\:\big| \phi_0(\vec{k}) \big|^2\:.  
\label{E0mom}
\eeq
This scalar product has the same units as the classical energy (see after~\eqref{E0}),
\beq \label{dimESP}
\big[ (\phi_0 | \psi_0) \big] = \ell^{-1}\:.
\eeq

Another scalar product can be obtained by identifying the classical solutions with
vectors of the one-particle Fock space: For given~$\phi_0$, we
seek a Schwartz function~$f$ such that
\[ \phi_0(x) = 2\, \re \,\la 0 \,|\, \hat{\phi}_0(x) \,|\, f \ket \:. \]
Using~\eqref{psidef} and the commutation relations, we find that~$f$ is uniquely determined by
\[ f(\hat{k}) = \phi_0(\vec{k}) \:. \]
We introduce the real scalar product~$\la .|. \ra$ on~$\H_0$ by~$\la \phi_0 | \phi_0 \ra := \la f | f \ra$.
Using~\eqref{psiscal}, this scalar product has the representation
\begin{align}
\la \phi_0 | \psi_0 \ra &= 
\frac{1}{(2 \pi)^3}\:\re \int \frac{d^3k}{2 \omega}\: \overline{\phi_0(\vec{k})} \: \psi_0(\vec{k})
\notag \\
&= \frac{1}{2 (2 \pi)^3}
\int \frac{d^3k}{2 \omega} \left(
\overline{\phi_0(\vec{k})} \,\psi_0(\vec{k}) + \phi_0(\vec{k}) \,\overline{\psi_0(\vec{k})} \right) .
\label{Fockclass}
\end{align}
This scalar product is Lorentz invariant. Moreover, comparing with~\eqref{E0mom} and~\eqref{dimESP},
one sees that it is dimensionless,
\beq \label{dimSP}
\big[ \la \phi_0 | \psi_0 \ra \big] = \ell^0\:.
\eeq
It is easy to verify that, up to a multiplicative constant, the scalar product~\eqref{Fockclass}
is indeed the only Lorentz invariant scalar product which can be introduced on the free classical scalar fields.
We thus obtain the isometric embedding
\beq \label{embed} \begin{split}
(\H_0, \la .|. \ra) &\hookrightarrow (\F, \la .|. \ra) \\
\phi_0 &\mapsto |f\ra \:.
\end{split}
\eeq

\subsection{Perturbative Solution of the Classical Cauchy Problem} \label{secnonlinear}
We now return to the nonlinear wave equation~\eqref{cfe}. For simplicity, we only consider
solutions which are smooth with spatially compactly support,
\[ \phi \in C^\infty(\R, C^\infty_0(\R^3)) \:. \]
We denote the set of all such solutions by~${\mathcal{H}}$.
Note that, since our equation is non-linear, ${\mathcal{H}}$ is not a vector space.
Moreover, there is no bilinear form or scalar product on~${\mathcal{H}}$. The only
quantity available is the classical energy~\eqref{E0},
\[ E \::\: {\mathcal{H}} \rightarrow \R^+_0\:. \]

For the description of the Cauchy problem, it is convenient to 
again combine~$\phi$ and its time derivative to a two-component function~$\Phi$,
\eqref{intval}, and to write the equation as a system of first order in time,
\beq \label{schroedinger}
\partial_t \Phi = \mathfrak{H}(\Phi) \quad \text{with} \quad
\mathfrak{H}(\Phi) = \begin{pmatrix} \dot{\phi} \\
\Delta \phi - \lambda \phi^3/6 \end{pmatrix} .
\eeq
Then the initial data at some time~$t_0$ is a vector
\[ \Phi(t_0) \in C^\infty_0(\R^3)^2\:. \]
The time evolution obtained by solving the Cauchy problem gives rise to the mapping
\[ U(t, t_0) \::\: C^\infty_0(\R^3)^2 \rightarrow C^\infty_0(\R^3)^2 \qquad \text{with} \qquad
\Phi(t) = U(t, t_0) \star \Phi(t_0)\:. \]
The operator~$U(t,t_0)$ is referred to as the {\em{time evolution operator}}.
Here the star emphasizes that it is a {\em{nonlinear}} operator.
But clearly, the time evolution has the group property,
\beq \label{group}
U(t'',t') \star U(t',t) = U(t'', t) \qquad \text{for all~$t,t',t'' \in \R$}\:.
\eeq

Let us rewrite the solution of the Cauchy problem perturbatively.
We decompose the operator~${\mathfrak{H}}$ into its linear
and nonlinear parts,
\[ \mathfrak{H}(\Phi) = \mathfrak{H}_0 \Phi + \lambda B(\Phi) \:, \]
where
\[ \mathfrak{H}_0 = \begin{pmatrix} 0 & 1 \\ \Delta & 0 \end{pmatrix} \:,\qquad
B(\Phi) = \frac{1}{6} \begin{pmatrix} 0 \\ -\phi^3 \end{pmatrix} . \]

In the case~$\lambda=0$ without interaction, the equation~\eqref{schroedinger}
is linear. The corresponding Cauchy problem can be solved formally by an exponential,
\beq \label{classfree}
\Phi(t) = e^{(t-t_0) \mathfrak{H}_0} \,\Phi(t_0)\:.
\eeq
In the next lemma, we express the linear time evolution operator~$e^{(t-t_0) \mathfrak{H}_0}$
in terms of the retarded Green's function~\eqref{Sret}.
\begin{Lemma} \label{lemmafree}
For any~$t \geq 0$, the operator~$e^{t \mathfrak{H}_0}$ can be written as
\beq \label{Rprop}
(e^{t \mathfrak{H}_0} \Phi)(\vec{x}) = \int R_t(\vec{x} - \vec{y})\: \Phi(\vec{y})\: d^3y\:,
\eeq
where the integral kernel is the distribution
\beq \label{Rtdef}
R_t(\vec{x}) =   \begin{pmatrix}
-\partial_t S^\wedge(t, \vec{x}) & -S^\wedge(t, \vec{x}) \\
-\Delta S^\wedge(t, \vec{x}) & -\partial_t S^\wedge(t, \vec{x})
\end{pmatrix} .
\eeq
\end{Lemma}
\Proof From the representation~\eqref{Sretrep} one sees that for positive times, $S^\wedge(t, \vec{x})$ is a solution of the wave equation. Thus a short calculation yields that the function~\eqref{Rprop} is
is a solution of the equation~$(\partial_t - {\mathfrak{H}}_0) (e^{t {\mathfrak{H}}_0} \Phi)=0$.
Next, one easily verifies from~\eqref{Sretrep} that
\[ \lim_{t \searrow 0} S^\wedge(t, \vec{x}) = 0 \qquad \text{and} \qquad
\lim_{t \searrow 0} \partial_t S^\wedge(t, \vec{x}) = -\delta^3(\vec{x})\:, \]
showing that~\eqref{Rprop} has the correct initial values at~$t=0$.
\QED

In order to treat the interaction perturbatively, we write~\eqref{schroedinger} as
\beq \label{schrP}
(\partial_t - \mathfrak{H}_0) \Phi =  \lambda B(\Phi) \:.
\eeq
Similar to the interaction picture, we set
\[ \Phi_\text{int}(t) = e^{-t \mathfrak{H}_0}\, \Phi(t) \:,\qquad
B_\text{int}(\Phi_\text{int}) = e^{-t \mathfrak{H}_0}\, B(e^{t \mathfrak{H}_0}\, \Phi_\text{int})\:. \]
Then~\eqref{schrP} simplifies to
\beq \label{intP}
\partial_t \Phi_\text{int} = \lambda B_\text{int}(\Phi_\text{int})\:.
\eeq
Making an ansatz as a formal expansion in powers of~$\lambda$,
\beq \label{phiintpert}
\Phi_\text{int}(t) = \Phi_\text{int}^{(0)} + \lambda \Phi_\text{int}^{(1)}(t) + \lambda^2 \Phi_\text{int}^{(2)}(t)
+ \cdots
\eeq
(where~$\Phi_\text{int}^{(0)}$ is a time independent wave with the correct initial data),
we obtain for~$n=1, 2, \ldots$ the equations
\beq \label{ODE}
\partial_t \Phi_\text{int}^{(n)}(x) = -\begin{pmatrix} 0 \\  \rho_\text{int}^{(n)}(x) \end{pmatrix}\:,
\eeq
where we set~$x=(t,\vec{x})$ and
\begin{align}
\begin{pmatrix} 0 \\  \rho_\text{int}^{(n)}(x) \end{pmatrix}
&= e^{-t \mathfrak{H}_0} \begin{pmatrix} 0 \\  \rho^{(n)}(x) \end{pmatrix} , \\
\rho^{(n)}(x) &= \sum_{a,b,c \text{ with } a+b+c=n-1} 
\frac{1}{6}\: \phi^{(a)}(x) \,\phi^{(b)}(x) \,\phi^{(c)}(x) \:. \label{rhondef}
\end{align}
Integrating~\eqref{ODE} on both sides and using
that~$\Phi_\text{int}^{(n)}(t_0,\vec{x})=0$ for~$n=1,2,\ldots$, we obtain
\[ \Phi_\text{int}^{(n)}(t) = -
\int_{t_0}^t \begin{pmatrix} 0 \\  \rho_\text{int}^{(n)}(\tau) \end{pmatrix} d\tau \:. \]
Transforming back to~$\Phi$ gives
\[ \Phi^{(n)}(t) = -
\int_{t_0}^t e^{(t-\tau) \mathfrak{H}_0} \begin{pmatrix} 0 \\  \rho^{(n)}(\tau) \end{pmatrix} d\tau \:. \]
Using the explicit form of the free time evolution operator~\eqref{Rtdef}, we obtain
\[  \Phi^{(n)}(t, \vec{x}) = -\int_{t_0}^t d\tau \int d^3y \:R_{t-\tau}(\vec{x}-\vec{y})
\begin{pmatrix} 0 \\  \rho^{(n)}(\tau, \vec{y})
\end{pmatrix} \:. \]

\begin{figure}
\begin{picture}(0,0)%
\includegraphics{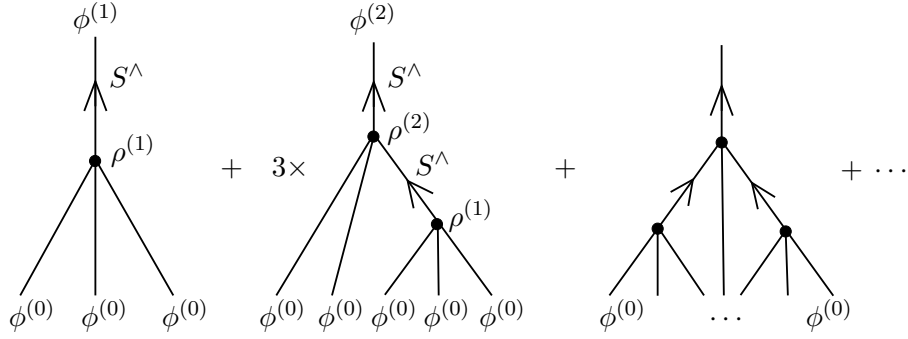}%
\end{picture}%
\setlength{\unitlength}{1533sp}%
\begingroup\makeatletter\ifx\SetFigFont\undefined%
\gdef\SetFigFont#1#2#3#4#5{%
  \reset@font\fontsize{#1}{#2pt}%
  \fontfamily{#3}\fontseries{#4}\fontshape{#5}%
  \selectfont}%
\fi\endgroup%
\begin{picture}(13475,5356)(2641,-4760)
\put(3691,209){\makebox(0,0)[lb]{\smash{{\SetFigFont{11}{13.2}{\rmdefault}{\mddefault}{\updefault}$\phi^{(1)}$}}}}
\put(8161,194){\makebox(0,0)[lb]{\smash{{\SetFigFont{11}{13.2}{\rmdefault}{\mddefault}{\updefault}$\phi^{(2)}$}}}}
\put(2656,-4606){\makebox(0,0)[lb]{\smash{{\SetFigFont{11}{13.2}{\rmdefault}{\mddefault}{\updefault}$\phi^{(0)}$}}}}
\put(3826,-4606){\makebox(0,0)[lb]{\smash{{\SetFigFont{11}{13.2}{\rmdefault}{\mddefault}{\updefault}$\phi^{(0)}$}}}}
\put(5131,-4606){\makebox(0,0)[lb]{\smash{{\SetFigFont{11}{13.2}{\rmdefault}{\mddefault}{\updefault}$\phi^{(0)}$}}}}
\put(6706,-4606){\makebox(0,0)[lb]{\smash{{\SetFigFont{11}{13.2}{\rmdefault}{\mddefault}{\updefault}$\phi^{(0)}$}}}}
\put(7651,-4606){\makebox(0,0)[lb]{\smash{{\SetFigFont{11}{13.2}{\rmdefault}{\mddefault}{\updefault}$\phi^{(0)}$}}}}
\put(8506,-4606){\makebox(0,0)[lb]{\smash{{\SetFigFont{11}{13.2}{\rmdefault}{\mddefault}{\updefault}$\phi^{(0)}$}}}}
\put(9361,-4606){\makebox(0,0)[lb]{\smash{{\SetFigFont{11}{13.2}{\rmdefault}{\mddefault}{\updefault}$\phi^{(0)}$}}}}
\put(10261,-4606){\makebox(0,0)[lb]{\smash{{\SetFigFont{11}{13.2}{\rmdefault}{\mddefault}{\updefault}$\phi^{(0)}$}}}}
\put(4276,-736){\makebox(0,0)[lb]{\smash{{\SetFigFont{11}{13.2}{\rmdefault}{\mddefault}{\updefault}$S^\wedge$}}}}
\put(8776,-736){\makebox(0,0)[lb]{\smash{{\SetFigFont{11}{13.2}{\rmdefault}{\mddefault}{\updefault}$S^\wedge$}}}}
\put(9226,-2161){\makebox(0,0)[lb]{\smash{{\SetFigFont{11}{13.2}{\rmdefault}{\mddefault}{\updefault}$S^\wedge$}}}}
\put(6076,-2176){\makebox(0,0)[lb]{\smash{{\SetFigFont{11}{13.2}{\rmdefault}{\mddefault}{\updefault}$+ \quad 3 \times$}}}}
\put(11476,-2176){\makebox(0,0)[lb]{\smash{{\SetFigFont{11}{13.2}{\rmdefault}{\mddefault}{\updefault}$+$}}}}
\put(9766,-2986){\makebox(0,0)[lb]{\smash{{\SetFigFont{11}{13.2}{\rmdefault}{\mddefault}{\updefault}$\rho^{(1)}$}}}}
\put(8776,-1576){\makebox(0,0)[lb]{\smash{{\SetFigFont{11}{13.2}{\rmdefault}{\mddefault}{\updefault}$\rho^{(2)}$}}}}
\put(4306,-1936){\makebox(0,0)[lb]{\smash{{\SetFigFont{11}{13.2}{\rmdefault}{\mddefault}{\updefault}$\rho^{(1)}$}}}}
\put(16101,-2216){\makebox(0,0)[lb]{\smash{{\SetFigFont{11}{13.2}{\rmdefault}{\mddefault}{\updefault}$+\: \cdots$}}}}
\put(12201,-4606){\makebox(0,0)[lb]{\smash{{\SetFigFont{11}{13.2}{\rmdefault}{\mddefault}{\updefault}$\phi^{(0)}\qquad \cdots \qquad \phi^{(0)}$}}}}
\end{picture}%
\caption{The tree diagrams in the perturbation expansion of the classical Cauchy problem.}
\label{figtree1}
\end{figure}

Setting~$y = (\tau, \vec{y})$, the perturbation expansion~\eqref{phiintpert} becomes
\beq \label{phipert}
\phi(x) = \sum_{n=0}^\infty \lambda^n \phi^{(n)} \:,
\eeq
where~$\phi^{(0)}=\phi_0$ is a solution of the free wave equation with the
correct initial values, and
\beq \label{iterate3}
\phi^{(n)}(x) =
\int_{\{ y^0 > t_0 \}} S^\wedge(x,y) \:\rho^{(n)}(y)\: d^4y \qquad (n \geq 1)\:.
\eeq
We thus obtain an expansion of~$\phi$ in terms of tree diagrams,
which can be depicted as shown in Figure~\ref{figtree1}.

Let us briefly compare the above perturbation expansion
for the classical time evolution with the perturbation expansion for the quantum field
as outlined in Section~\ref{secpertqu}: The equation for the classical evolution equation
in the interaction picture~\eqref{intP} can be regarded as the analog of the
quantum time evolution operator~$U$ in~\eqref{ans2}. Although formally similar, they differ
in that the classical evolution is non-linear, whereas the quantum evolution
is linear.
The resulting perturbation expansion for the classical field differs from that for the
quantum field in that only non-loop diagrams appear (see Figures~\ref{figloop1} and~\ref{figtree1}).
Moreover, the diagrams corresponding to the perturbative expansion of the classical field always involve the retarded Green's function, instead of the Feynman propagator, and they all have exactly one outgoing leg.

\subsection{A Nonlinear Classical Measurement Process} \label{secmeasure}
Let us try to mimic the construction in linear scattering theory as outlined in Section~\ref{sec21}
(see~\eqref{lindyn}--\eqref{Sclass}).
Denoting the free classical dynamics~\eqref{classfree} by~$U_0(t,t') = e^{(t-t') {\mathfrak{H}}_0}$,
we can introduce in analogy to~\eqref{Omegapm}
\[ \Omega_\pm \psi := U(t_0, \pm T) \star U_0(\pm T, t_0) \,\psi \::\: \H_0 \rightarrow {\mathcal{H}} \:, \]
where at time~$T$ we identify free and interacting solutions.
Note that the operators~$\Omega_\pm$ are nonlinear. Moreover, as~${\mathcal{H}}$ is not a linear
space, we cannot take their adjoints. Rewriting~\eqref{Sclass} with the inverse, the obvious idea
is to define the nonlinear scattering operator by
\[ (\Omega_-)^{-1} \star \Omega_+ = U_0(t_0, -T) \star U(-T, T) \star U_0( T, t_0) \:, \]
where in the last step we used the group property~\eqref{group}.
As the operator~$U_0(T, t_0)$ maps a free solution at time~$t_0$ to the same solution at time~$T$,
it is the identity on~$\H_0$. Thus the naive ansatz for the scattering operator is
\[ S \::\: \H_0 \rightarrow \H_0 \:, \qquad
S \,\phi_\text{out} = U(-T, T) \star \phi_\text{out} \:. \]
Similar to~\eqref{matrix}, a matrix element of the scattering operator would be given by
\beq \label{matrixtry}
( \beta_\text{out} , \alpha_\text{in} ) = ( S^{-1} \star \beta_\text{in} , \alpha_\text{in} )
\eeq
(alternatively, one could work with the scalar product~\eqref{Fockclass}; this would make no
difference for the following consideration).
However, this naive approach does not work for the following reasons.
The first problem is that the construction manifestly distinguishes a direction of time.
This can be seen from the fact that, similar to~\eqref{iterate3}, the
perturbation expansion for~$U(-T, T)$ will involve only advanced but no retarded
Green's functions. Moreover, as~$U(-T, T)$ is a nonlinear operator,
the expression~\eqref{matrixtry} is nonlinear in~$\beta_\text{in}$, but it is linear
in~$\alpha_\text{in}$. A related problem is that~$S$ is not unitary,
\[ ( S^{-1} \star \beta_\text{in}, \alpha_\text{in} ) \neq 
( \beta_\text{in}, S \star \alpha_\text{in} ) \:. \]
This inequality is obvious because the left side is linear in~$\alpha_\text{in}$, whereas the
right side is not.

In view of these problems, it is not obvious conceptually how to introduce a non\-li\-near scattering
operator. In order to clarify what we have in mind, we first discuss the physical setting in more detail.
As explained at the beginning of Section~\ref{sec3}, we 
consider a solution~$\phi$ of the nonlinear wave equation~\eqref{cfe} which
as~$t \rightarrow \pm \infty$ goes over asymptotically
to free solutions~$\phi_\text{in}$ and~$\phi_\text{out}$ (see Figure~\ref{figscatter}).
Our wave is classical. First of all, this means that the field is not a quantum field in the
sense of second quantization. Moreover, the usual point of view is that a classical field
can be determined at every space-time point
by measuring its effect on a point-like ``test particle'', which satisfies the classical equations of motion.
By making this test particle sufficiently small, one can measure the classical field
to any precision without changing the system
(note that this is different in a quantum system, where a measurement changes the system
by projecting the quantum state to an eigenstate of the observable).
In particular, one can determine~$\phi$ at some initial time~$t_0$
and seek for a solution at later times by solving the Cauchy problem (see Section~\ref{sec21}).
The unique solvability of the Cauchy problem shows that the classical system is
deterministic. Moreover, the finite propagation speed shows that causality is respected.
In a perturbative treatment, causality becomes manifest by the fact that only
retarded Green's functions appear (see Section~\ref{secnonlinear}).

In what follows, we modify the above setting in that we do not want to refer to test particles.
This is because our model only involves the bosonic field~$\phi$, but there are no
classical point particles which could be used for measurements.
Moreover, we do not want to make the idealized assumption that
the system can be measured without disturbing it.
More specifically, the only observables which we want to use for measurements
are {\em{differences}} of {\em{classical energies}}.
Working with the classical energy seems natural because it is a conserved quantity
of the system. We restrict attention to energy differences because a constant offset to
the classical energy can be interpreted as being the contribution
of a background field which cannot be detected.
With these restricted measurements, the observer cannot determine the field at every space-time point.
In particular, he cannot determine the field~$\phi$ pointwise at
an initial time~$t_0$. Therefore, he cannot solve the Cauchy problem to make
predictions on the future. 
Thus, although our system is still classical and deterministic, the observer 
cannot predict the outcome of an experiment with certainty.

We now discuss how a scattering process can be described in this restricted framework where
only differences of classical energies are allowed for measurements.
Suppose that an observer at some large time~$T$ wants to detect the result of the scattering process.
The most obvious method to make observations is to
modify the wave~$\phi$ by some ``test wave'' $\delta \phi$ and to consider
how the classical energy~\eqref{E0}
changes. If the test wave is taken into account linearly, the energy is perturbed by
\[ \delta E = \int_{t=T} 
\Big( \dot{\phi} \:\delta \dot{\phi} + \vec{\nabla} \phi \cdot \vec{\nabla} (\delta \phi) + \frac{\lambda}{6}\:
\phi^3\: \delta \phi \Big)\, d^3x \:. \]
Using that at time~$T$, the outgoing wave packets should be so spread out that the
term involving~$\lambda$ can be dropped, we obtain the simple expression
\beq \label{perturblinear}
\delta E = ( \phi \,,\, \delta \phi )\:,
\eeq
where~$( .,. )$ is the scalar product obtained by polarizing the free energy~\eqref{E0free}
at time~$T$.
Despite the fact that we are considering purely classical fields, the
resulting situation has similarity with a measurement process in quantum mechanics.
Namely, the computation of the expectation value in~\eqref{perturblinear} can be
interpreted that a measurement of the wave~$\phi$ is performed
with a prepared ``end-state'' $\delta \phi$. By modifying~$\delta \phi$, one can determine~$\phi$
completely. In particular, one can measure the distribution of~$\phi$ in momentum space.

\begin{figure}
\begin{picture}(0,0)%
\includegraphics{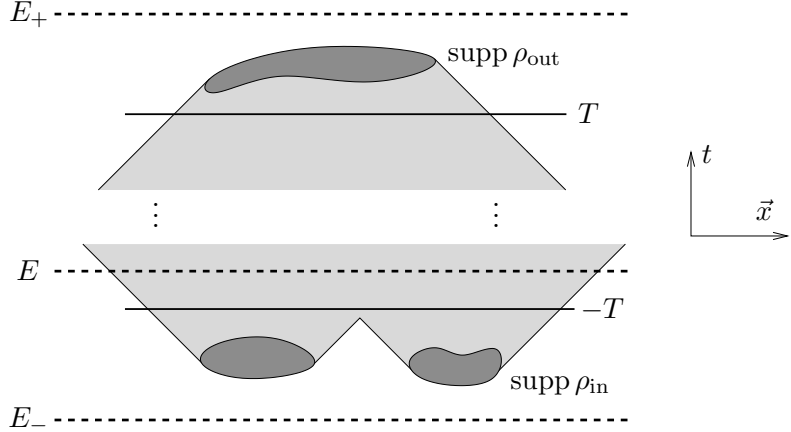}%
\end{picture}%
\setlength{\unitlength}{1492sp}%
\begingroup\makeatletter\ifx\SetFigFont\undefined%
\gdef\SetFigFont#1#2#3#4#5{%
  \reset@font\fontsize{#1}{#2pt}%
  \fontfamily{#3}\fontseries{#4}\fontshape{#5}%
  \selectfont}%
\fi\endgroup%
\begin{picture}(12997,7242)(7771,-6382)
\put(15826,-2926){\makebox(0,0)[lb]{\smash{{\SetFigFont{11}{13.2}{\rmdefault}{\mddefault}{\updefault}$\vdots$}}}}
\put(10171,-2926){\makebox(0,0)[lb]{\smash{{\SetFigFont{11}{13.2}{\rmdefault}{\mddefault}{\updefault}$\vdots$}}}}
\put(17296,-4441){\makebox(0,0)[lb]{\smash{{\SetFigFont{11}{13.2}{\rmdefault}{\mddefault}{\updefault}$-T$}}}}
\put(17251,-1201){\makebox(0,0)[lb]{\smash{{\SetFigFont{11}{13.2}{\rmdefault}{\mddefault}{\updefault}$T$}}}}
\put(19306,-1861){\makebox(0,0)[lb]{\smash{{\SetFigFont{11}{13.2}{\rmdefault}{\mddefault}{\updefault}$t$}}}}
\put(20206,-2806){\makebox(0,0)[lb]{\smash{{\SetFigFont{11}{13.2}{\rmdefault}{\mddefault}{\updefault}$\vec{x}$}}}}
\put(15091,-166){\makebox(0,0)[lb]{\smash{{\SetFigFont{11}{13.2}{\rmdefault}{\mddefault}{\updefault}$\text{supp} \,\rho_\text{out}$}}}}
\put(16126,-5611){\makebox(0,0)[lb]{\smash{{\SetFigFont{11}{13.2}{\rmdefault}{\mddefault}{\updefault}$\text{supp} \,\rho_\text{in}$}}}}
\put(7786,-6241){\makebox(0,0)[lb]{\smash{{\SetFigFont{11}{13.2}{\rmdefault}{\mddefault}{\updefault}$E_-$}}}}
\put(7921,-3811){\makebox(0,0)[lb]{\smash{{\SetFigFont{11}{13.2}{\rmdefault}{\mddefault}{\updefault}$E$}}}}
\put(7801,509){\makebox(0,0)[lb]{\smash{{\SetFigFont{11}{13.2}{\rmdefault}{\mddefault}{\updefault}$E_+$}}}}
\end{picture}%
\caption{A nonlinear classical measurement process.}
\label{fig1}
\end{figure}

The important point for what follows is that the relation~\eqref{perturblinear} only holds in the
linear approximation. If the amplitude of~$\delta \phi$ is increased, it has a nonlinear effect on~$\phi$,
which influences the result of the measurement process.
In order to analyze such nonlinear effects, we consider the situation shown in Figure~\ref{fig1}.
It is convenient to describe the incoming field
by an inhomogeneity~$\rho_\text{in}$ which lies in the past of the interaction region and
generates a wave moving to the future. Similarly, the observer is described by an
inhomogeneity~$\rho_\text{out}$, which is located in the future of the interaction region and
generates a wave moving to the past. The resulting wave~$\phi$ will be a solution of the
nonlinear inhomogeneous wave equation
\beq \label{phiinhom}
-\Box \phi - \frac{\lambda}{6}\: \phi^3 = \frac{1}{2}\left( \rho_\text{in} + \rho_\text{out} \right) .
\eeq
Due to the inhomogeneities, the classical energy~\eqref{E0} is not conserved.
But it is conserved in the regions of space-time where both~$\rho_\text{in}$ and~$\rho_\text{out}$
vanish. We can thus distinguish three different energies: the energy~$E$ in the interaction region, the
energy~$E_-$ in the past of~$\rho_\text{in}$, and the energy~$E_+$ in the future of~$\rho_\text{out}$.

As explained above, we take the point of view that the observer can only measure energy differences.
Moreover, as we are interested in the effect of the interaction, it is convenient to
also subtract the corresponding energies of the free fields, which we denote by
the additional superscript ``free''. This leads us to introduce the quantity
\beq \label{DelEdef}
\Delta E := (E-E^\text{free}) - (E_+ - E_+^\text{free})\:.
\eeq
It gives the energy shift caused by the interaction as being
detected by the observer at late times. We interpret~$\Delta E$ as the result of our classical
measurement process.

\subsection{The Global Perturbation Expansion} \label{sec32}
In what follows, similar to~\eqref{notfirst} we use the short notations
\[ S^\wedge(x, \rho) \equiv \int S^\wedge(x,y)\: \rho(y)\: d^4y\:,\quad
S^\wedge(\rho_1, \rho_2) \equiv \iint S^\wedge(x,y)\: \rho_1(x)\, \rho_2(y)\:
d^4x\, d^4y\:, \]
also analogously for other two-point distributions.
In order to model the classical measurement process (see~\eqref{phiinhom} and Figure~\ref{fig1}),
we choose the free solution~$\phi^{(0)}$ as
\beq \label{phi0def}
\begin{split}
\phi^{(0)} &= \phi_\text{in} + \phi_\text{out} \qquad \text{with} \\
\phi_\text{in}(x) := \frac{1}{2}\: S^\wedge(x, \,&\rho_\text{in}) \:,\qquad
\phi_\text{out}(x) := \frac{1}{2}\:S^\vee(x, \rho_\text{out}) \:.
\end{split}
\eeq
It is a solution of the inhomogeneous linear equation~$-\Box \phi_0 = (\rho_\text{in} + \rho_\text{out})/2$.
In order to construct a corresponding solution of the nonlinear equation~\eqref{phiinhom},
we make a perturbation ansatz as a formal power series in~$\lambda$,
\beq \label{pser}
\phi = \sum_{n=0}^\infty \lambda^n \,\phi^{(n)} \:.
\eeq
By direct calculation one verifies that the functions~$\phi^{(1)}, \phi^{(2)}, \ldots$ can be written iteratively as
\beq \label{perturb}
\begin{split}
\phi^{(n)}(x) &\:= \int S^{(n)}(x,y) \: \rho^{(n)}(y)\: d^4y \qquad \text{with} \\
\rho^{(n)}(y) &:= \chi_I \!\!\!\!\!\! \sum_{a,b,c \text{ with } a+b+c={n-1}}
\frac{1}{6}\:\big(\phi^{(a)}\, \phi^{(b)}\, \phi^{(c)} \big) (y)\:,
\end{split}
\eeq
where~$S^{(n)}$ is a general Green's function, meaning that it is a distributional solution of the equation
\[ -\Box S^{(n)}(x) = \delta^4(x)\:. \]
Here~$\chi_I$ is the characteristic function defined by~$\chi_I(x)=1$ if~$x \in I$
and~$\chi_I(x)=0$ otherwise. It has the effect that in~\eqref{perturb} we only integrate over
the interaction region. This is a technical simplification which will be used in the proof
of Lemma~\ref{lemmaEK}. It has no physical significance because in the scattering process
under consideration, we assumed that the dynamics outside the interaction region is linear,
so that cubic expressions in~$\phi$ can be omitted.

Since we want~$\phi$ to be a real-valued function,
the Green's function should be real. This gives rise to the general ansatz
\beq \label{Snrep}
S^{(n)}(x,y) = S_0(x,y) + c_n \, P_0(x,y) + i d_n\, K_0(x,y)\:,
\eeq
where~$S_0$ is the causal Green's function
\beq
S_0(x,y) = \frac{1}{2} \, \big( S^\vee + S^\wedge \big)(x,y)
= \int \frac{d^4k}{(2 \pi)^4}\: \frac{\text{PP}}{k^2}\: e^{-ik(x-y)} \:, \label{S0def}
\eeq
whereas~$K_0, P_0$ are the fundamental
solutions of the free wave equation~\eqref{K0def} and
\beq P_0(x,y) = \int \frac{d^4k}{(2 \pi)^4}\: \delta(k^2)\: e^{-ik(x-y)} \:, \label{P0def}
\eeq
and the constants~$c_n$ and~$d_n$ are free real parameters
(note that~$P_0$ is real-valued, whereas~$K_0$ is purely imaginary).
Thus the perturbation expansion involves two real parameters, which can be
chosen freely to every order in perturbation theory.

At this stage, there seems no reason why the parameters~$c_n$ and~$d_n$ should
depend on the order~$n$ of the perturbation expansion (note that
this is different for the perturbation expansion for fermions, where the concept of the
Dirac sea and normalization conditions give rise to a non-trivial combinatorics of the operator
products~\cite{sea, grotz}).
Therefore, in the remainder of this paper we always make the simplest choice
where all parameters~$c_n$ and~$d_n$ vanish,
\beq \label{Snrel}
S^{(n)}(x,y) = S_0(x,y) \:.
\eeq
The resulting perturbation expansion~\eqref{perturb}
and~\eqref{Snrel} is neither advanced nor retarded and should thus be regarded as
a global expansion in space-time. Therefore, we refer to it as the {\em{global
perturbation expansion}}. It gives rise to a nonlinear mapping~$\p$
which to every~$\phi^{(0)} \in \H^{(0)}$ associates a corresponding solution~$\phi$ of the nonlinear equation,
\beq \label{pdef}
\p \::\: \H^{(0)} \rightarrow {\mathcal{H}}\:.
\eeq

We close this section with a few remarks.
We first point out that our choice of Green's functions~\eqref{Snrel} was mainly a matter
of simplicity and convenience.
But more general choices of the coefficients~$c_n$ and~$d_n$ are also possible and might be worth
considering in the future.
We also remark that the vanishing of the coefficients~$d_n$ in~\eqref{Snrep} can be motivated by
time reflection symmetry as follows: Note that in our classical scattering process
we did not distinguish a direction of time, meaning that the setting is
symmetric under the transformations
\beq \label{timesymm}
t \rightarrow -t \qquad \text{and} \qquad \phi_\text{in} \leftrightarrow \phi_\text{out}
\eeq
(clearly, this transformation exchanges the observer in the future with an observer in the past,
and also leads to the replacement~$E_+ \leftrightarrow E_-$).
Therefore, it seems natural to impose that also the perturbation expansion should
be invariant under~\eqref{timesymm}.
Since the distributions~$S_0$ and~$P_0$
are symmetric in its two arguments, whereas~$K_0$ is anti-symmetric, this leads us to impose
that~$d_n=0$.

We finally note that the Feynman propagator~\eqref{Feynprop} can be written as
\[ \triangle_F = S_0 - i \pi P_0 \:. \]
This is of the form~\eqref{Snrep}, but with the parameter~$c_n=-i \pi$ being imaginary.
This shows once more that the Feynman propagator is complex-valued (indeed, it is imaginary in spacelike
directions). Therefore, it cannot be used for the perturbation expansion of real-valued classical fields.

\subsection{A Perturbation Expansion of the Classical Energy} \label{secEperturb}
We now consider the classical energy~\eqref{E0} of our solution~$\phi$,
\beq \label{Erel}
E(t) = \int_{t=\text{const}} 
\left( \frac{1}{2}\: \dot{\phi}^2 + \frac{1}{2}\: |\nabla \phi|^2 + \frac{\lambda}{4!}\:
\phi^4 \right) d^3x \:.
\eeq
As in Figure~\ref{fig1}, we denote the energy in the interaction region by~$E$,
whereas~$E_\pm$ are the energies for large positive and negative times, respectively.

\begin{Prp} \label{lemmaEclass2} The classical energies~$E$ and~$E_\pm$ have the expansions
\begin{align}
E_+ &= \frac{i \pi}{4} \dot{K}_0 \big(\rho_\text{in} + \tilde{\rho},\:
\rho_\text{in} + \tilde{\rho} \big) 
\:+\: \lim_{t \rightarrow \infty} \frac{\lambda}{4!} \int_{t=\text{\rm{const}}} \phi^4\, d^3x
\label{Ep} \\
E &= \frac{i \pi}{4} \dot{K}_0 \big(\rho_\text{in} - \rho_\text{out} + \tilde{\rho},\:
\rho_\text{in} - \rho_\text{out} + \tilde{\rho} \big)
\:+\: \frac{\lambda}{4!} \int_{t=T} \phi^4\, d^3x \label{Eclass2} \\
&= \frac{i \pi}{4} \dot{K}_0 \big(\rho_\text{in} - \rho_\text{out} - \tilde{\rho},\:
\rho_\text{in} - \rho_\text{out} - \tilde{\rho} \big)
\:+\: \frac{\lambda}{4!} \int_{t=-T} \phi^4\, d^3x \label{Eclass3} \\
E_- &= \frac{i \pi}{4} \dot{K}_0 \big(\rho_\text{out} + \tilde{\rho},\:
\rho_\text{out} + \tilde{\rho} \big) 
\:+\: \lim_{t \rightarrow -\infty} \frac{\lambda}{4!} \int_{t=\text{\rm{const}}} \phi^4\, d^3x
\:, \label{Em}
\end{align}
where~$\dot{K}_0(x,y) = \partial_{x^0} K_0(x,y)$ is the time derivative of the
distribution~\eqref{K0def}.
Here the function~$\phi$ is defined inductively by~\eqref{phi0def}--\eqref{perturb},
and~$\tilde{\rho}$ is given by
\beq \label{rhotdef}
\tilde{\rho} = \sum_{n=1}^\infty \rho^{(n)} \:.
\eeq
\end{Prp}
\Proof Due to energy conservation, the energy~$E$ can be computed at any
time~$t \in [-T, T]$. We first compute it at time~$t=T$ to derive~\eqref{Eclass2}.
We again polarize the free energy to obtain the scalar product~$(.,.)$
introduced in~\eqref{E0free}.
We now substitute the perturbation expansion~\eqref{pser} into~\eqref{Erel}.
Multiplying out the free part of the energy, we need to compute expressions of the form
\[ \big( S^\vee(.,x) ,\, S_0(.,y) \big) \:,\quad
\big( S^\wedge(.,x) ,\, S_0(.,y) \big) \:,\quad \big( S_0(.,x) ,\, S_0(.,y) \big)\:, \ldots \:. \]
Furthermore, we know that the argument~$x$ of the factors~$S^\wedge(.,x)$
and~$S_0(.,x)$ always lies to the past of time~$T$ (because~$x \in I \cup \supp \rho_\text{in}$),
whereas the argument of the factors~$S^\vee(.,x)$ lies to the future of the time~$T$
(because in this case, $x \in \supp \rho_\text{out})$).
According to~\eqref{Krep}, \eqref{Sretrep}, \eqref{S0def} and~\eqref{K0def}, we may thus
exchange the Green's functions with the replacement rules
\beq \label{reps}
S^\wedge(.,x) \rightarrow -2 \pi i \: K_0(.,x) \:, \quad
S^\vee(.,x) \rightarrow 2 \pi i\: K_0(.,x) \:, \quad
S_0(.,x) \rightarrow -i\pi\: K_0(.,x)
\eeq
by corresponding fundamental solutions~$K_0$. After these replacements, we can apply
Lemma~\ref{lemmaEK} below. In particular, we obtain for any~$n, n' \geq 1$,
\begin{align*}
&\big( \phi^{(n)} ,\, \phi^{(n')} \big) = (-i \pi)^2
\iint \big( K_0(.,x) ,\, K_0(., y) \big)\: \rho^{(n)}(x)\: \rho^{(n')}(y)
\:d^4x\, d^4y \\
&\quad= -\pi^2 \left(- \frac{i}{4 \pi} \right)
\iint \dot{K}_0(x,y)\: \rho^{(n)}(x)\:
\rho^{(n')}(y) \:d^4x\, d^4y
= \frac{i \pi}{4}\: \dot{K}_0 \big( \rho^{(n)}, \rho^{(n')} \big) \:.
\end{align*}
Adding all the terms of the perturbation expansion gives~\eqref{Eclass2}.

To derive the expansion~\eqref{Eclass3}, we compute the energy
similarly at time~$t=-T$. Then the interaction region lies in the future of~$t$,
and therefore the replacement rule for~$S_0$ in~\eqref{reps} is to be modified to
\[ S_0(.,x) \rightarrow i\pi\, K_0(.,x) \:. \]

The expansion for~$E_+$ is derived similar to~\eqref{Eclass2}.
However, as the argument~$x$ of the factors~$S^\vee(.,x)$ now lies in the past of~$t$,
all the term involving a scalar product with~$S^\vee(.,x)$ drop out.
Then energy conservation allows us to take the limit~$t \rightarrow \infty$.
Similarly, the expansion for~$E_-$ is obtained from~\eqref{Eclass3}
by omitting the terms involving~$S^\wedge(.,x)$
and taking the limit~$t \rightarrow -\infty$.
\QED
It remains compute the energy scalar product of two factors~$K_0$.
\begin{Lemma} \label{lemmaEK}
The following relation holds in the sense of distributions,
\begin{align*}
\big( K_0(., x) ,\, K_0(. , y) \big) &= -\frac{i}{4 \pi} \dot{K}_0(y,x) \:.
\end{align*}
\end{Lemma}
\Proof
We write the scalar product in momentum space as
\begin{align*}
\big( K_0(., x) ,\, K_0(. , y) \big)
&= \lim_{\tau \rightarrow \infty} \: \frac{1}{2} \int d^3z
\int \frac{d^4k}{(2 \pi)^4}\: \epsilon(k^0)\:\delta(k^2)\: e^{-ik^0 (\tau-x^0)
+ i \vec{k}(\vec{z}-\vec{x})} \\
&\quad\; \times
\int \frac{d^4q}{(2 \pi)^4}\: \epsilon(q^0)\:\delta(q^2)\: e^{-iq^0 (\tau-y^0) + i \vec{q}(\vec{z}-\vec{y})}
\left(- k^0 q^0 -  \vec{k} \vec{q} \right) .
\end{align*}
Carrying out the spatial integral, we only get a contribution if~$\vec{q} = -\vec{k}$.
Setting~$\Omega=q^0$, we thus obtain
\begin{align*}
\big( K_0(., x) ,\, K_0(. , y) \big)
&\:= \lim_{\tau \rightarrow \infty} \: \frac{1}{4 \pi} \int \frac{d^4k}{(2 \pi)^4} \int_{-\infty}^\infty d\Omega\:
\epsilon(k^0)\:\delta(k^2)\:  \epsilon(\Omega)\: \delta(\Omega^2 - |\vec{k}|^2)\\
&\qquad \times e^{-ik^0 (\tau-x^0) - i \Omega (\tau-y^0)+ i \vec{k}(\vec{y}-\vec{x})}
\left(- k^0 \Omega + |\vec{k}|^2 \right) .
\end{align*}
Due to the two delta distributions, the integrand vanishes unless~$\Omega = \pm k^0$.
In the case~$\Omega = +k^0$, the last factor vanishes, because
\[ - k^0 \Omega + |\vec{k}|^2 = (k^0)^2 - |\vec{k}|^2 = k^2 = 0 \:. \]
Thus we may set~$\Omega = -k^0$ to obtain
\begin{align*}
\big( K_0(., x) ,\, K_0(. , y) \big)
&\:= -\frac{1}{4 \pi} \int \frac{d^4k}{(2 \pi)^4} \:\frac{1}{2 |\vec{k}|}\:
\delta(k^2)\: e^{-ik^0 (y^0 - x^0)+ i \vec{k}(\vec{y}-\vec{x})}\:
2 |\vec{k}|^2 \\
&\:= -\frac{1}{4 \pi} \int \frac{d^4k}{(2 \pi)^4} \: \epsilon(\omega)\: \omega\:
\delta(k^2)\: e^{-ik (y-x)} \:.
\end{align*}
Comparing with~\eqref{K0def} gives the result.
\QED

We now analyze the result of Proposition~\ref{lemmaEclass2}.
First, if we assume that the solution~$\phi$ 
is dissipative and for large times goes over to a solution of the free wave equation,
the last integrals in~\eqref{Ep} and~\eqref{Em} vanish.
This dissipation could be made precise using
the techniques described in~\cite{strauss, ginibre+velo, roach, strauss2}.
Here we take the decay property for granted and simply drop the last integrals
in~\eqref{Ep} and~\eqref{Em}. Similarly, if we choose the interaction time~$T$ sufficiently large,
we can also drop the last integrals in~\eqref{Eclass2} and~\eqref{Eclass3}.
Then taking the difference of~\eqref{Eclass2} and~\eqref{Eclass3} gives the identity
\beq \label{K0id}
\dot{K}_0 \big(\rho_\text{in} - \rho_\text{out},\: \tilde{\rho} \big) = 0 \:.
\eeq
Using this identity, we obtain
\begin{align*}
E_+ &= \frac{i \pi}{4} \dot{K}_0 \big(\rho_\text{in} + \tilde{\rho},\:
\rho_\text{in} + \tilde{\rho} \big) \\
E &= \frac{i \pi}{4} \dot{K}_0 \big(\rho_\text{in} - \rho_\text{out},\:
\rho_\text{in} - \rho_\text{out}\big)
+ \frac{i \pi}{4} \dot{K}_0 \big(\tilde{\rho}, \tilde{\rho} \big) \\
E_- &= \frac{i \pi}{4} \dot{K}_0 \big(\rho_\text{out} + \tilde{\rho},\:
\rho_\text{out} + \tilde{\rho} \big) .
\end{align*}
Here the function~$\tilde{\rho}$ describes the component of the wave generated by
the interaction, whereas~$\rho_\text{in}$ and~$\rho_\text{out}$ describe the free
incoming and outgoing components. In order to analyze how the energies are effected
by the interaction, we subtract the free energies as given by
\begin{align*}
E_+^\text{free} &= \frac{i \pi}{4} \dot{K}_0 \big(\rho_\text{in}, \rho_\text{in} \big) \\
E^\text{free} &= \frac{i \pi}{4} \dot{K}_0 \big(\rho_\text{in} - \rho_\text{out},
\rho_\text{in} - \rho_\text{out}\big) \\
E_-^\text{free} &= \frac{i \pi}{4} \dot{K}_0 \big(\rho_\text{out} , \rho_\text{out} \big) .
\end{align*}
We thus obtain the compact formulas
\begin{align*}
E - E^\text{free} &= \frac{i \pi}{4} \dot{K}_0 \big(\tilde{\rho}, \tilde{\rho} \big) \\
E_+ - E_+^\text{free} &= E_- - E_-^\text{free} = \frac{i \pi}{2} \dot{K}_0 \big(\rho_\text{out}, \tilde{\rho} \big) 
+ \frac{i \pi}{4} \dot{K}_0 \big(\tilde{\rho}, \tilde{\rho} \big) .
\end{align*}

Let us discuss these formulas. First, the formulas for the free energies
can be understood immediately from the fact that for large positive (negative) times, only the
wave generated by~$\rho_\text{in}$ (resp.\ $\rho_\text{out}$) is present, 
whereas in the interaction region, both waves are superimposed.
The interaction affects the energies by two different contributions: The
term
\[ \frac{i \pi}{4} \dot{K}_0 \big(\tilde{\rho}, \tilde{\rho} \big) \]
is delocalized in space-time in the sense that it adds to the energies~$E_\pm$
and~$E$ in exactly the same way. As a consequence, this contribution is not
accessible to the observer who can only measure the energy
difference~$E - E_+$, whereas a constant offset to the energy is not detectable.
The situation is different for the contribution
\[ \frac{i \pi}{2} \dot{K}_0 \big(\rho_\text{out}, \tilde{\rho} \big) , \]
which only effects the energy in the interaction region. The observer at time~$T$
can detect this contribution by measuring the classical field energy before and
after modifying the field by~$\rho_\text{out}$. We now identify this contribution
with~\eqref{DelEdef} and bring it into a more convenient form.

\begin{Prp} In the classical measurement process described in Section~\ref{secmeasure},
the energy shift~\eqref{DelEdef} as measured by the observer at time~$T$ is given by
\beq \label{Idef}
\Delta E = \frac{1}{4}\: \dot{S}^\vee(\tilde{\rho}, \rho_\text{\rm{out}}) = 
-\frac{1}{4}\: \dot{S}^\wedge(\tilde{\rho}, \rho_\text{\rm{in}}) \:.
\eeq
\end{Prp}
\Proof Comparing the above formulas for the classical energies with~\eqref{DelEdef}, we find
\[ \Delta E = \frac{i \pi}{2} \dot{K}_0 \big(\rho_\text{out}, \tilde{\rho} \big)
= \frac{i \pi}{2} \dot{K}_0 \big(\rho_\text{in}, \tilde{\rho} \big) . \]
The function~$\tilde{\rho}$ is localized in the interaction region. Therefore, we can
use~\eqref{reps} to replace~$\dot{K}$ by the time derivatives of the causal Green's functions.
\QED
We remark that~\eqref{Idef} shows that our constructions are symmetric under
time reversals (note that a time reversal corresponds to the replacements~$\rho_\text{out}
\leftrightarrow \rho_\text{in}$, $S^\vee \leftrightarrow S^\wedge$ and~$\partial_t \leftrightarrow -\partial_t$).

\begin{figure}
\begin{picture}(0,0)%
\includegraphics{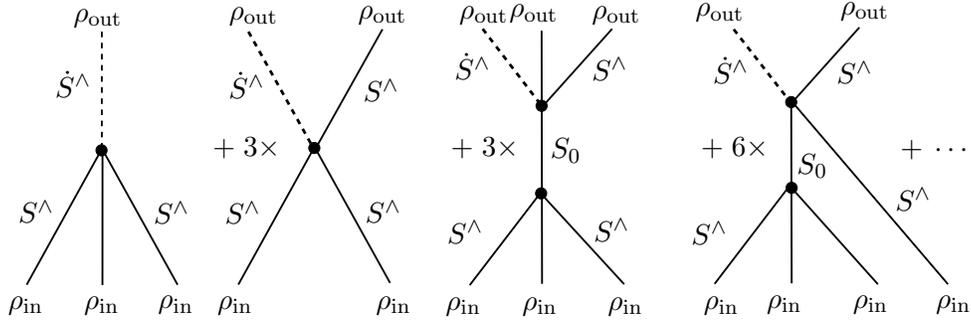}%
\end{picture}%
\setlength{\unitlength}{1533sp}%
\begingroup\makeatletter\ifx\SetFigFont\undefined%
\gdef\SetFigFont#1#2#3#4#5{%
  \reset@font\fontsize{#1}{#2pt}%
  \fontfamily{#3}\fontseries{#4}\fontshape{#5}%
  \selectfont}%
\fi\endgroup%
\begin{picture}(15236,5302)(2528,-4667)
\put(3596,149){\makebox(0,0)[lb]{\smash{{\SetFigFont{11}{13.2}{\rmdefault}{\mddefault}{\updefault}$\rho_\text{out}$}}}}
\put(2543,-4501){\makebox(0,0)[lb]{\smash{{\SetFigFont{11}{13.2}{\rmdefault}{\mddefault}{\updefault}$\rho_\text{in}$}}}}
\put(3766,-4494){\makebox(0,0)[lb]{\smash{{\SetFigFont{11}{13.2}{\rmdefault}{\mddefault}{\updefault}$\rho_\text{in}$}}}}
\put(4965,-4493){\makebox(0,0)[lb]{\smash{{\SetFigFont{11}{13.2}{\rmdefault}{\mddefault}{\updefault}$\rho_\text{in}$}}}}
\put(5925,-4501){\makebox(0,0)[lb]{\smash{{\SetFigFont{11}{13.2}{\rmdefault}{\mddefault}{\updefault}$\rho_\text{in}$}}}}
\put(15294,-2307){\makebox(0,0)[lb]{\smash{{\SetFigFont{11}{13.2}{\rmdefault}{\mddefault}{\updefault}$S_0$}}}}
\put(9836,201){\makebox(0,0)[lb]{\smash{{\SetFigFont{11}{13.2}{\rmdefault}{\mddefault}{\updefault}$\rho_\text{out}$}}}}
\put(10646,224){\makebox(0,0)[lb]{\smash{{\SetFigFont{11}{13.2}{\rmdefault}{\mddefault}{\updefault}$\rho_\text{out}$}}}}
\put(11981,201){\makebox(0,0)[lb]{\smash{{\SetFigFont{11}{13.2}{\rmdefault}{\mddefault}{\updefault}$\rho_\text{out}$}}}}
\put(13653,238){\makebox(0,0)[lb]{\smash{{\SetFigFont{11}{13.2}{\rmdefault}{\mddefault}{\updefault}$\rho_\text{out}$}}}}
\put(15993,260){\makebox(0,0)[lb]{\smash{{\SetFigFont{11}{13.2}{\rmdefault}{\mddefault}{\updefault}$\rho_\text{out}$}}}}
\put(9614,-4516){\makebox(0,0)[lb]{\smash{{\SetFigFont{11}{13.2}{\rmdefault}{\mddefault}{\updefault}$\rho_\text{in}$}}}}
\put(13784,-4509){\makebox(0,0)[lb]{\smash{{\SetFigFont{11}{13.2}{\rmdefault}{\mddefault}{\updefault}$\rho_\text{in}$}}}}
\put(13593,-3456){\makebox(0,0)[lb]{\smash{{\SetFigFont{11}{13.2}{\rmdefault}{\mddefault}{\updefault}$S^\wedge$}}}}
\put(2703,-3104){\makebox(0,0)[lb]{\smash{{\SetFigFont{11}{13.2}{\rmdefault}{\mddefault}{\updefault}$S^\wedge$}}}}
\put(4893,-3119){\makebox(0,0)[lb]{\smash{{\SetFigFont{11}{13.2}{\rmdefault}{\mddefault}{\updefault}$S^\wedge$}}}}
\put(8320,-3110){\makebox(0,0)[lb]{\smash{{\SetFigFont{11}{13.2}{\rmdefault}{\mddefault}{\updefault}$S^\wedge$}}}}
\put(3304,-1076){\makebox(0,0)[lb]{\smash{{\SetFigFont{11}{13.2}{\rmdefault}{\mddefault}{\updefault}$\dot{S}^\wedge$}}}}
\put(6108,-1091){\makebox(0,0)[lb]{\smash{{\SetFigFont{11}{13.2}{\rmdefault}{\mddefault}{\updefault}$\dot{S}^\wedge$}}}}
\put(8282,-1145){\makebox(0,0)[lb]{\smash{{\SetFigFont{11}{13.2}{\rmdefault}{\mddefault}{\updefault}$S^\wedge$}}}}
\put(9704,-2026){\makebox(0,0)[lb]{\smash{{\SetFigFont{11}{13.2}{\rmdefault}{\mddefault}{\updefault}$+ \; 3 \times$}}}}
\put(11312,-2066){\makebox(0,0)[lb]{\smash{{\SetFigFont{11}{13.2}{\rmdefault}{\mddefault}{\updefault}$S_0$}}}}
\put(12010,-3403){\makebox(0,0)[lb]{\smash{{\SetFigFont{11}{13.2}{\rmdefault}{\mddefault}{\updefault}$S^\wedge$}}}}
\put(6116,179){\makebox(0,0)[lb]{\smash{{\SetFigFont{11}{13.2}{\rmdefault}{\mddefault}{\updefault}$\rho_\text{out}$}}}}
\put(8201,195){\makebox(0,0)[lb]{\smash{{\SetFigFont{11}{13.2}{\rmdefault}{\mddefault}{\updefault}$\rho_\text{out}$}}}}
\put(9640,-3440){\makebox(0,0)[lb]{\smash{{\SetFigFont{11}{13.2}{\rmdefault}{\mddefault}{\updefault}$S^\wedge$}}}}
\put(6048,-3119){\makebox(0,0)[lb]{\smash{{\SetFigFont{11}{13.2}{\rmdefault}{\mddefault}{\updefault}$S^\wedge$}}}}
\put(9759,-790){\makebox(0,0)[lb]{\smash{{\SetFigFont{11}{13.2}{\rmdefault}{\mddefault}{\updefault}$\dot{S}^\wedge$}}}}
\put(11980,-800){\makebox(0,0)[lb]{\smash{{\SetFigFont{11}{13.2}{\rmdefault}{\mddefault}{\updefault}$S^\wedge$}}}}
\put(16885,-2909){\makebox(0,0)[lb]{\smash{{\SetFigFont{11}{13.2}{\rmdefault}{\mddefault}{\updefault}$S^\wedge$}}}}
\put(15941,-854){\makebox(0,0)[lb]{\smash{{\SetFigFont{11}{13.2}{\rmdefault}{\mddefault}{\updefault}$S^\wedge$}}}}
\put(13937,-851){\makebox(0,0)[lb]{\smash{{\SetFigFont{11}{13.2}{\rmdefault}{\mddefault}{\updefault}$\dot{S}^\wedge$}}}}
\put(16970,-2037){\makebox(0,0)[lb]{\smash{{\SetFigFont{11}{13.2}{\rmdefault}{\mddefault}{\updefault}$+\: \cdots$}}}}
\put(5849,-2025){\makebox(0,0)[lb]{\smash{{\SetFigFont{11}{13.2}{\rmdefault}{\mddefault}{\updefault}$+ \; 3 \times$}}}}
\put(14804,-4479){\makebox(0,0)[lb]{\smash{{\SetFigFont{11}{13.2}{\rmdefault}{\mddefault}{\updefault}$\rho_\text{in}$}}}}
\put(16199,-4479){\makebox(0,0)[lb]{\smash{{\SetFigFont{11}{13.2}{\rmdefault}{\mddefault}{\updefault}$\rho_\text{in}$}}}}
\put(8490,-4501){\makebox(0,0)[lb]{\smash{{\SetFigFont{11}{13.2}{\rmdefault}{\mddefault}{\updefault}$\rho_\text{in}$}}}}
\put(10852,-4516){\makebox(0,0)[lb]{\smash{{\SetFigFont{11}{13.2}{\rmdefault}{\mddefault}{\updefault}$\rho_\text{in}$}}}}
\put(12239,-4517){\makebox(0,0)[lb]{\smash{{\SetFigFont{11}{13.2}{\rmdefault}{\mddefault}{\updefault}$\rho_\text{in}$}}}}
\put(17542,-4480){\makebox(0,0)[lb]{\smash{{\SetFigFont{11}{13.2}{\rmdefault}{\mddefault}{\updefault}$\rho_\text{in}$}}}}
\put(13747,-2025){\makebox(0,0)[lb]{\smash{{\SetFigFont{11}{13.2}{\rmdefault}{\mddefault}{\updefault}$+ \; 6 \times$}}}}
\end{picture}%
\caption{The perturbation expansion of
the energy shift~$\Delta E$ in the classical measurement process.}
\label{figtree2}
\end{figure}

Expressing~$\tilde{\rho}$ by~\eqref{rhotdef} and the perturbation expansion~\eqref{phi0def}--\eqref{perturb},
we obtain an expansion of~$\Delta E$ in terms of tree diagrams. The resulting diagrams are
shown in Figure~\ref{figtree2}.
This perturbation expansion differs from that for the solution of the Cauchy problem 
(see Figure~\ref{figtree1}) in several ways: First, we can now have several outgoing legs.
Second, the inner lines now involve the Green's function~$S_0$ (instead of the
retarded Green's function~$S^\wedge$); this is a direct consequence of our choice
of the Green's function~\eqref{Snrel} in the global perturbation expansion.
Finally, the time derivative in~\eqref{Idef} has the effect that in
each diagram, exactly one of the outgoing legs is differentiated with respect to~$t$.
This leg is denoted by a dotted line. This time derivative obviously destroys the
Lorentz invariance. This can be understood from the fact that as an energy,
$\Delta E$ is not a Lorentz scalar but the zero component of a Lorentz vector.

\subsection{The Classical Scattering Operator and Classical $n$-Point Functions} \label{seccscatter}
Let us reconsider the previous constructions.
The result of the classical measurement process was associated to
a suitable difference of classical energies~$\Delta E$ as given by \eqref{DelEdef}.
To compute these energies, we started from the free solution~\eqref{phi0def},
performed the global perturbation expansion, and could then evaluate
the energy integrals (see Proposition~\ref{lemmaEclass2}).
Describing the perturbation expansion by the nonlinear operation~\eqref{pdef}, we thus
obtain the nonlinear mapping
\beq \label{Efrak}
\Delta E \::\: \H_0 \times \H_0 \rightarrow \R \:,\qquad
\Delta E(\phi_\text{out}, \phi_\text{in}) := \Delta E(\p \star (\phi_\text{out}+ \phi_\text{in}))\:.
\eeq
Plugging in the perturbation expansion~\eqref{perturb} and multiplying out, we get an infinite sum of
multilinear mappings,
\beq \label{multi}
\Delta E(\phi_\text{out}, \phi_\text{in}) = \sum_{p,q=0}^\infty \Delta E_{p,q}
(\underbrace{\phi_\text{out},\ldots, \phi_\text{out}}_{\text{$p$ arguments}},
\underbrace{\phi_\text{in}, \ldots, \phi_\text{in}}_{\text{$q$ arguments}}) \:.
\eeq

At this point, it is helpful to 
apply the standard procedure of rewriting the multilinear operators as linear operators
on the tensor product. To this end, we introduce the free $p$-particle classic bosonic Fock space
\[ \F_\text{class}^n = (\underbrace{\H_0 \otimes \cdots \otimes \H_0}_{\text{$n$ factors}})_\text{symm} \:, \]
where ``symm'' means total symmetrization (and~$\F^0$ are simply the real numbers).
On~$\F^n_\text{class}$ we let~$\la .|. \ra$ be the scalar product induced
from the Lorentz invariant scalar product~\eqref{Fockclass} on~$\H_0$.
Taking the direct sum gives the
\[ \text{\em{classical Fock space}} \qquad \F_\text{class} = \bigoplus_{n=0}^\infty \F^n_\text{class} \:. \]
We remark that lifting the isometric embedding~\eqref{embed} to the tensor product yields a
canonical isometric embedding of the Fock spaces,
\[ (\F_\text{class}, \la .|. \ra) \hookrightarrow (\F, \la .|. \ra) \:. \]
The two Fock spaces differ by the fact that~$\F_\text{class}$ is a real, whereas~$\F$ is a complex
vector space. In fact, $\F$ is the complexification of the image of~$\F_\text{class}$ under the above
embedding.

As the multilinear mappings in~\eqref{multi} give rise to unique linear mappings on the
corresponding symmetric tensor products, we obtain a unique bilinear operator
\beq \label{Efrakbil}
\Delta E \::\: \F_\text{class} \times \F_\text{class} \rightarrow \R \:.
\eeq
The classical scattering operator~$S$ is introduced by
\beq \label{Sclass2}
S \,:\, \F_\text{class} \rightarrow \F_\text{class} \qquad \text{with} \qquad
\Delta E(\Phi_\text{out}, \Phi_\text{in}) = \frac{1}{2}
\la \partial_t \Phi_\text{out} \,|\, S^\dagger \, \Phi_\text{in} \ra \:.
\eeq
In the next lemma we bring the classical scattering operator into a more
explicit form. Our proof will also explain why the factor~$\partial_t$ in~\eqref{Sclass2}
is needed in order to make the definition Lorentz invariant.

\begin{Prp} \label{prpscatter} The scattering operator has the matrix elements
\beq \label{Sndef}
\bra \phi^1_\text{\rm{out}} \otimes \cdots \otimes \phi^p_\text{\rm{out}}
\,|\, S^\dagger \, \phi^1_\text{\rm{in}} \otimes \cdots \otimes \phi^q_\text{\rm{in}} \ket
= S_{p+q}(\phi^1_\text{\rm{out}}, \ldots, \phi^p_\text{\rm{out}}, 
\phi^1_\text{\rm{in}}, \ldots, \phi^q_\text{\rm{in}} )\:.
\eeq
Here the functions~$S_{p+q}$ have a perturbation expansion in terms of Feynman tree
diagrams. The order~$k$ and the total number of lines~$N$ are given by~\eqref{Nrel}
and~\eqref{treerel}, where~$n=p+q$.
The contributions to~$S_n$ without the outer lines are given analytically by
\beq \label{contri2}
\lambda^k \int d^4y_1\: \cdots \int d^4y_k\;
\prod_{\ell=1}^{N-n} \:S_0(.,.)
\eeq
(where~$S_0$ is the Green's function~\eqref{S0def}).
\end{Prp} \noindent
The functions~$S_n(x_1, \ldots, x_n)$ are referred to as the {\bf{classical {\em{n}}-point functions}}.
\Proof[Proof of Proposition~\ref{prpscatter}]
Combining~\eqref{Idef} with~\eqref{phi0def}, we obtain
\beq \label{DelEdot}
\Delta E = \frac{1}{2} \int \tilde{\rho}(x)\: \dot{\phi}_\text{out}(x)\: d^4x\:.
\eeq
Expressing~$\tilde{\rho}$ by the perturbation series~\eqref{rhotdef} and~\eqref{perturb}, one gets
an expansion in terms of tree diagrams. This explains~\eqref{Nrel} and~\eqref{treerel}.
In order to get hold of the combinatorics,
we consider the summand~$\Delta E_{p,q}$ in~\eqref{multi} and number the
arguments by~$\phi_\text{out}^1, \ldots, \phi_\text{out}^p$ and~$\phi_\text{in}^1, \ldots, \phi_\text{in}^q$.
Moreover, we restrict attention to the contributions to~$\Delta E_{p,q}$ corresponding to a
fixed tree diagram, whose outer lines are labeled by the outer arguments.
According to~\eqref{DelEdot}, exactly one of the outer arguments~$\phi_\text{in}^1, \ldots,
\phi_\text{in}^p$ involves a time derivative. If we fix this dotted outer leg, the only combinatorial
factor comes from the freedom to permute the three incoming legs at each vertex.
This leads to a factor~$3!$ at each vertex, which cancels the factor~$1/6$ in~\eqref{perturb}.
We thus obtain
\begin{align}
&\Delta E_{p,q}(\phi_\text{out}^1, \ldots, \phi_\text{out}^p,
\phi_\text{in}^1, \ldots, \phi_\text{in}^q) \notag \\
&= \frac{1}{2} \left( S_n( \dot{\phi}_\text{out}^1, \ldots, \phi_\text{out}^p,
\phi_\text{in}^1, \ldots, \phi_\text{in}^q)
+ \cdots + S_n( \phi_\text{out}^1, \ldots, \dot{\phi}_\text{out}^p,
\phi_\text{in}^1, \ldots, \phi_\text{in}^q) \right) \notag \\
&= \frac{1}{2}\: \bra \partial_t (\phi^1_\text{\rm{out}} \otimes \cdots \otimes \phi^p_\text{\rm{out}})
\,|\, S^\dagger \, \phi^1_\text{\rm{in}} \otimes \cdots \otimes \phi^q_\text{\rm{in}} \ket\:,
\label{DelErel}
\end{align}
where~$S_n$ is formed according to~\eqref{contri2}, and~$S^\dagger$ is given by~\eqref{Sndef}.
Comparing with~\eqref{Sclass2} shows that~$S^\dagger$ indeed agrees with the
classical scattering operator. This concludes the proof.
\QED

Let us compare the classical $n$-point functions with the $n$-point functions of
quantum field theory as computed in Lemma~\ref{lemmacombi}.
One obvious difference is that the classical $n$-point functions involve {\em{only tree
diagrams}}, whereas the loop diagrams of quantum field theory are missing.
Before comparing the combinatorics~\eqref{contri2} with~\eqref{contri},
one should note that in~\eqref{Sndef} we inserted free fields~$\phi_\text{in}^\ell$
and~$\phi_\text{out}^\ell$ as arguments. This is why in~\eqref{contri2} there are
no propagators for the outer lines (this corresponds to the usual procedure
of considering the ``truncated'' diagrams). This difference also accounts for the
additional factors of~$i$ in~\eqref{contri}. Namely, according to~\eqref{Nrel}
and~\eqref{treerel}, for tree diagrams the identity~$n=N-k-1$ holds, and thus
the factor~$i^{N-k}$ in~\eqref{contri} can be generated in the classical
scattering operator if to every outer line we associate a factor of~$i$.
Then the combinatorial factors in~\eqref{contri2} coincide with those in~\eqref{contri},
up to an irrelevant prefactor which could be absorbed into the definition of the
classical scattering operator~\eqref{Sclass2} (the reason for the convention~\eqref{Sclass2}
is that we want the classical scattering operator to be real).
We conclude that on the level of tree diagrams, the classical and the quantum scattering
agree, including the combinatorics of the expansion.
The only difference is that in the classical $n$-point functions, the inner lines are
described by the Green's function~$S_0$, whereas in the quantum scattering process
the Feynman propagator~$\triangle_F$ appears. This difference can be understood from the
fact that the classical field is real, and therefore its dynamics cannot be described with the
complex-valued Feynman propagator (see the end of Section~\ref{sec32}).

We note that the classical scattering operator also differs from the quantum scattering operator
in that it only involves the connected diagrams. As a consequence, the classical scattering
operator~$S : \F_\text{class} \rightarrow \F_\text{class}$ is not unitary (this is obvious
already without interaction, because in the case~$\lambda=0$, the operator~$S$ is the identity on the
one-particle subspace~$\F^1_\text{class}$ but vanishes on the many-particle subspaces).
A possible strategy to obtain a unitary classical scattering operator would be to
take the sum of tensor products of~$S$, like for example the expression
\[ \sum_{n=0}^\infty \frac{1}{n!} \: \underbrace{S \otimes \cdots \otimes S}_{\text{$n$ factors}} \:. \]
This would give a unitary operator in the case~$\lambda=0$. However, it is not clear whether this
operator would also be unitary in the case with interaction. If not, one could take its polar decomposition
and absorb the non-unitary part into the scalar product of the classical Fock spaces.
Entering the details of this analysis or discussing its potential physical significance
goes beyond the scope of the present paper.

We finally remark how the identity~\eqref{K0id} can be understood in terms of Feynman
diagrams. In the proof of Proposition~\ref{prpscatter} we computed~$\Delta E$
with the first equation in~\eqref{Idef}. The identity~\eqref{K0id} makes it possible to
compute~$\Delta E$ alternatively by the second equation in~\eqref{Idef}.
In analogy to~\eqref{DelEdot} and~\eqref{DelErel}, we thus obtain
\begin{align}
\Delta E &= -\frac{1}{2} \int \tilde{\rho}(x)\: \dot{\phi}_\text{in}(x)\: d^4x \\
\Delta E_{p,q} &= -\frac{1}{2}\: \bra \phi^1_\text{\rm{out}} \otimes \cdots \otimes \phi^p_\text{\rm{out}}
\,|\, S^\dagger \, \partial_t (\phi^1_\text{\rm{in}} \otimes \cdots \otimes \phi^q_\text{\rm{in}}) \ket
\end{align}
Subtracting the last equation from~\eqref{DelErel}, we find that~\eqref{K0id} is equivalent to the
identity for the $S$-matrix elements
\[ \la \partial_t \Phi_\text{out} \,|\, S^\dagger \, \Phi_\text{in} \ra
+ \la \Phi_\text{out} \,|\, S^\dagger \, \partial_t \Phi_\text{in} \ra = 0 \:. \]
This means that the value of the Feynman diagram remains unchanged if both the
incoming and the outgoing waves are translated in time.
It corresponds to the usual concept of {\em{conservation of quantum mechanical energy}}
in a Feynman diagram, stating that the incoming energy, defined as the sum of the frequencies of the
incoming waves, coincides with the outgoing energy.

\section{Nonlinear Classical Scattering in a Stochastic Background Field} \label{sec4}

\subsection{A Free Stochastic Field} \label{sec41}
We begin by analyzing a free stochastic field~$\xi$.
By ``free'' we mean that~$\xi \in \H_0$ is a solution of the linear wave equation.
Similar to~\eqref{intval}, we can characterize~$\xi$ by its initial values for example at some fixed time~$t_0$,
\[ \Xi := (\xi, \dot{\xi})|_{t_0} \in C^\infty_0(\R^3) \times C^\infty_0(\R^3)\:. \]
``Stochastic'' means that we prescribe a probability distribution for~$\xi$
which we assume to be {\em{Gaussian}}. We write the corresponding probability measure~$\D \Xi$
on the initial data as
\beq \label{gausspos}
\D \Xi = \frac{1}{Z}\: e^{-\frac{1}{2}\: F(\Xi, \Xi)}\: d\xi \wedge d\dot{\xi} \:,
\eeq
where~$Z$ is a normalization constant, and~$F$ is a positive bilinear form, i.e.\
\[ F(\Xi, \Xi) = \int \big\la \Xi(\vec{x}), F \,\Xi(\vec{x}) \big\ra_{\R^2}\: d^3x \]
with~$F$ being a linear operator on the initial data. Next, it is natural to assume that our probability
measure is {\em{translation invariant}}, meaning that the operator~$F$ is diagonal in momentum space.
Thus setting
\beq \label{Xik}
\Xi(\vec{x}) = \int \frac{d^3k}{(2 \pi)^3}\: {\Xi}(\vec{k})\: e^{i \vec{k} \vec{x}} \:,
\eeq
the operator~$F$ can be written as
\[ \widehat{(F \Xi)}(\vec{k}) = {F}(\vec{k}) \:{\Xi}(\vec{k}) \:, \]
where~${F}(\vec{k})$ is a~$2 \times 2$-matrix.
The fact that~$\Xi$ is real-valued and that~$F$ acts on a space of real-valued functions implies that
\beq \label{realcond}
{\Xi}(-\vec{k}) = \overline{{\Xi}(\vec{k})} \qquad \text{and} \qquad
{F}(-\vec{k}) = \overline{{F}(\vec{k})}\:,
\eeq
where the bar denotes complex conjugation of each vector or matrix component.
Using Plancherel's theorem, we can write the bilinear form~$F$ as
\beq \label{Fmom}
F(\Xi, \Xi) = \int \frac{d^3k}{(2 \pi)^3}\:
\big\la {\Xi}(\vec{k}), {F}(\vec{k}) \,{\Xi}(\vec{k}) \big\ra_{\C^2} \:.
\eeq

A Gaussian probability measure is completely characterized by the {\em{covariance}}~${\mathcal{C}}$
defined as the expectation value of a quadratic polynomial, i.e.\ in our setting
\beq \label{defcovariance}
{\mathcal{C}}(x,y) := \int \xi(x)\: \xi(y)\: \D \Xi \:.
\eeq
The expectation values of higher order polynomials can then be computed
using Wick's theorem (similar as outlined in Section~\ref{secpertqu} in the context
of path integrals or see for example~\cite{glimm+jaffe}).

\begin{Lemma} \label{lemmacov}
For any space-time points~$x, y \in I$, the covariance~\eqref{defcovariance} is given by
\begin{align}
{\mathcal{C}}(x,y) =\:& \int \frac{d^4k}{(2 \pi)^4} \: \delta(k^2)\: \frac{\pi}{|\omega|} \:
e^{-i \vec{k} (\vec{x} - \vec{y})} \notag \\
& \times \bigg[ e^{i \omega (x^0-y^0)}\:
\Big\la \begin{pmatrix} -i \omega \\ 1 \end{pmatrix}
, {F}(\vec{k})^{-1} \begin{pmatrix} -i \omega \\ 1 \end{pmatrix} \Big\ra_{\C^2} \label{term1} \\
& \quad\quad - e^{i \omega (x^0 + y^0 + 2t)} \:
\Big\la \begin{pmatrix} -i \omega \\ 1 \end{pmatrix}
, {F}(\vec{k})^{-1} \begin{pmatrix} i \omega \\ 1 \end{pmatrix} \Big\ra_{\C^2} \bigg] . \label{term2}
\end{align}
\end{Lemma}
\Proof In momentum space, the Gaussian measure~\eqref{gausspos} becomes
\beq \label{gaussmom}
\D \Xi = \frac{1}{\tilde{Z}}\: \exp \left( -\frac{1}{2} \int \frac{d^3 k}{(2 \pi)^3}\:\la {\Xi}(\vec{k}), {F}(\vec{k}) \,{\Xi}(\vec{k}) \ra_{\C^2} \right)
d{\Xi}_1(\vec{k}) \wedge d{\Xi}_2(\vec{k}) \:.
\eeq
By the standard calculation rules for Gaussian measures (see for example~\cite{glimm+jaffe}),
we obtain
\beq \label{gaussdir}
\int {\Xi}(\vec{k}) \otimes {\Xi}(\vec{k}')^\dagger \: \D \Xi = (2 \pi)^3\: \delta^3(\vec{k} - \vec{k}')\:
{F}(\vec{k})^{-1}\:.
\eeq

Next, we know from~\eqref{Rprop} that
\[ \xi(x) = \int
\Big( \begin{matrix}
-\dot{S}_0(x, (t, \vec{y})) & -S_0(x, (t, \vec{y}))
\end{matrix} \Big)
\: \Xi(\vec{y})\: d^3y \]
and thus
\[ \xi(x)\: \xi(y) = \int_{\R^3 \times \R^3} \!\!\!\!\!\!\! d^3z_1\: d^3z_2\:
\Big\la \begin{pmatrix} -\dot{S}^0(x, (t, \vec{z}_1)) \\ -S^0(x, (t, \vec{z}_1)) \end{pmatrix}
, \Xi(\vec{z}_1) \otimes \Xi(\vec{z}_2)^\dagger
\begin{pmatrix} -\dot{S}^0(y, (t, \vec{z}_2)) \\ -S^0(y, (t, \vec{z}_2)) \end{pmatrix} \Big\ra_{\C^2} \]
Similar to~\eqref{reps}, we may apply the replacement rule
\[ S_0^\wedge(x,y) \rightarrow -2 \pi i\: K_0(x,y) \:. \]
As in Lemma~\ref{lemmaEK}, for the momenta we use the notation~$k = (\omega, \vec{k})$
and~$q = (\Omega, \vec{q})$. We then obtain
\begin{align*}
\xi(x)\: \xi(y) = 4 \pi^2 & \int \frac{d^4k}{(2 \pi)^4} \:e^{ik x}\:\epsilon(\omega)\: \delta(k^2) 
\int \frac{d^4q}{(2 \pi)^4} \:e^{-iq y}\: \epsilon(\Omega)\: \delta(q^2) \\
&\times\: e^{i (\omega - \Omega) t}\:
\Big\la \begin{pmatrix} i \omega \\ -1 \end{pmatrix}
, {\Xi}(\vec{k}) \otimes {\Xi}(\vec{q})^\dagger
\begin{pmatrix} i \Omega \\ -1 \end{pmatrix} \Big\ra_{\C^2}\:.
\end{align*}
Now we can carry out the Gaussian integral with~\eqref{gaussdir},
\begin{align*}
\int \xi(x)\: \xi(y)\: \D \Xi &= 4 \pi^2 \int \frac{d^4k}{(2 \pi)^4} \:e^{ik x}\:\epsilon(\omega)\: \delta(k^2) 
\int \frac{d\Omega}{2 \pi} \:e^{-i \Omega y^0 + i \vec{k} \vec{y}}\: \epsilon(\Omega)\: \delta(\Omega - |\vec{k}|^2) \\
&\qquad\qquad \times\: e^{i (\omega - \Omega) t}\:
\Big\la \begin{pmatrix} i \omega \\ -1 \end{pmatrix}
, {F}(\vec{k})^{-1} \begin{pmatrix} i \Omega \\ -1 \end{pmatrix} \Big\ra_{\C^2} \\
&= \pi \int \frac{d^4k}{(2 \pi)^4} \:e^{ik x}\:\epsilon(\omega)\: \delta(k^2) \\
&\qquad \times\: \frac{1}{\Omega}\: e^{-i \Omega y^0 + i \vec{k} \vec{y}}\: e^{i (\omega - \Omega) t}\:
\Big\la \begin{pmatrix} i \omega \\ -1 \end{pmatrix}
, {F}(\vec{k})^{-1} \begin{pmatrix} i \Omega \\ -1 \end{pmatrix} \Big\ra_{\C^2}
\bigg|_{\Omega= \pm \omega} \:.
\end{align*}
Rearranging this expression gives the result.
\QED \noindent
Let us discuss the result of this lemma. We first point out that the covariance is obviously
real-valued and symmetric under permutations and~$x$ and~$y$. Moreover, it satisfies the free wave
equation,
\[ {\mathcal{C}}(x,y) = {\mathcal{C}}(y,x) \qquad \text{and} \qquad
\Box_x \,{\mathcal{C}}(x,y) = 0 \:. \]
The last equation is obvious from~\eqref{defcovariance} because the factors~$\xi(x)$ and~$\xi(y)$
are solutions of the wave equation. Alternatively, it can also be verified in the expression
of Lemma~\ref{lemmacov} by applying the wave operator and using the factor~$\delta(k^2)$
in the integrand on the right. In particular, it is {\em{impossible to arrange}}
that~${\mathcal{C}}(x,y)$ is {\em{a Green's function}}.

In order to obtain a covariant theory, we would like the covariance~${\mathcal{C}}$ to be Lorentz
invariant. This raises the questions for which choices of~$F$ the covariance is Lorentz invariant,
and which Lorentz invariant distributions~${\mathcal{C}}$ can be realized. We first explain what the naive
answer to these questions is, and then prove that this naive guess is indeed correct.
The general Lorentz invariant bi-solution of the scalar wave equation is a linear combination
of the distributions~$P_0$ and~$K_0$ introduced in~\eqref{P0def} and~\eqref{K0def}.
As~$K_0(x,y)$ is anti-symmetric under permutations of~$x$ and~$y$ and~$P_0(x,y)$ is real-valued,
we conclude that~${\mathcal{C}}$ should be a real multiple of~$P_0(x,y)$.
\begin{Prp} Assume that the covariance~\eqref{defcovariance} of the Gaussian measure~\eqref{gausspos}
is Lorentz invariant. Then there is a parameter~$\beta>0$ such that
\begin{align}
{\mathcal{C}}(x,y) &= \frac{1}{\beta}\: P_0(x,y) \label{Cid} \\
F(\Xi, \Xi) &= \frac{\beta}{\pi}\: \la \xi | \xi \ra \:, \label{QuE}
\end{align}
where~$\la \xi | \xi \ra$ denotes the scalar product~\eqref{Fockclass}.
\end{Prp}
\Proof Lorentz invariance clearly implies that~${\mathcal{C}}(x,y)$ must be rotationally symmetric,
meaning that~${F}(\vec{k})$ must depend only on~$|\vec{k}|$. Combining this fact with
the second equation in~\eqref{realcond}, we conclude that~${F}(\vec{k})$ must have real matrix entries.
Thus its inverse can be written as
\beq \label{Fform}
{F}(\vec{k})^{-1} = \begin{pmatrix} a & b \\ b & c \end{pmatrix} .
\eeq
with real functions~$a, b, c$ depending on~$|\vec{k}|$. Next, Lorentz invariance implies that~${\mathcal{C}}(x,y)$
should be independent of~$F$, meaning that the contribution~\eqref{term2} must vanish, i.e.
\[ \Big\la \begin{pmatrix} -i \omega \\ 1 \end{pmatrix}
, {F}(\vec{k})^{-1} \begin{pmatrix} i \omega \\ 1 \end{pmatrix} \Big\ra_{\C^2} = 0 
\qquad \text{for~$\omega = \pm |\vec{k}|$} \:. \]
Combining this condition with~\eqref{Fform} we conclude that~$b=0$ and that~$c = \omega^2 a$.
Then the expectation value in~\eqref{term1} becomes
\[ \Big\la \begin{pmatrix} -i \omega \\ 1 \end{pmatrix}
, {F}(\vec{k})^{-1} \begin{pmatrix} -i \omega \\ 1 \end{pmatrix} \Big\ra_{\C^2} = 
2 \omega^2\, a(|\vec{k}|)\:. \]
Substituting these results into the formula of Lemma~\ref{lemmacov}, we obtain
\[ {\mathcal{C}}(x,y) = \int \frac{d^4k}{(2 \pi)^3} \: \delta(k^2)\: e^{i k(x - y)} \:|\omega|\: a(|\vec{k}|) . \]
This distribution is Lorentz covariant only if~$a(|\vec{k}|) = (2 \pi \beta\, |\vec{k}|)^{-1}$ for a
suitable real parameter~$\beta$. This gives~\eqref{Cid}. Moreover,
\beq \label{Fmom2}
{F}(\vec{k})^{-1} = \frac{1}{2 \pi \beta\, |\vec{k}|} \begin{pmatrix} 1 & 0 \\ 0 & \omega^2 \end{pmatrix} 
\qquad \text{and thus} \qquad
{F}(\vec{k})= 2 \pi \beta \begin{pmatrix} |\vec{k}| & 0 \\ 0 & 1/|\vec{k}| \end{pmatrix} \:.
\eeq

Next, we write~$\xi$ 
similar to~\eqref{Fourierrep} as a Fourier integral over the upper and lower mass cone, i.e. 
\[ \xi(x) = \frac{1}{(2 \pi)^4} \int \frac{d^3k}{2 \omega}
\left( \xi(\vec{k}) \:e^{-i \omega t + i \vec{k} \vec{x}}
+ \overline{\xi(\vec{k})} \:e^{i \omega t - i \vec{k} \vec{x}} \right) \:. \]
Taking the time derivative, evaluating for simplicity at~$t=0$ and comparing with~\eqref{Xik}, we find that
\[ {\Xi}_1(\vec{k}) = \frac{1}{4 \pi \omega} \left( {\xi}(\vec{k}) + \overline{{\xi}(-\vec{k})} \right)
\qquad \text{and} \qquad
{\Xi}_2(\vec{k}) = - \frac{i}{4 \pi} \left( {\xi}(\vec{k}) - \overline{{\xi}(-\vec{k})} \right) . \]
Hence the functional~$F$ as given by~\eqref{Fmom} and~\eqref{Fmom2} becomes
\beq
F(\Xi, \Xi) = \frac{\beta}{2 \pi} \int \frac{d^3k}{(2 \pi)^3}\: \frac{1}{\omega}
\:\big| {\xi}(\vec{k}) \big|^2 \:.
\eeq
Comparing with~\eqref{Fockclass} gives the result.
\QED

Recall from~\eqref{dimSP} that the scalar product~$\la \xi | \xi \ra$ is dimensionless.
Moreover, the functional~$F$, being the argument of the exponential in~\eqref{gausspos},
must clearly be dimensionless. Hence the parameter~$\beta$ in~\eqref{QuE}
is also dimensionless,
\beq \label{betadim}
[\beta] = \ell^0\:.
\eeq
This parameter describes the strength of stochastic background field.
In the limit~$\beta \rightarrow \infty$, the covariance~\eqref{Cid} vanishes, so that
no stochastic field is present. If~$\beta$ is decreased, the amplitude of the stochastic
field gets larger.

\subsection{The Perturbation Expansion and Classical Scattering} \label{sec42}
We now want to include the stochastic background field in the interaction.
To this end, we first add the free stochastic field~$\xi$
to the free in- and outgoing fields in~\eqref{phi0def},
\beq \label{phi0tot}
\phi_0 = \phi_\text{in} + \phi_\text{out} + \xi \:.
\eeq
The corresponding interacting field is then introduced exactly as in Section~\ref{sec32}
by the perturbation series~\eqref{pser} and~\eqref{perturb}.
Expanding the classical energy exactly as in Section~\ref{secEperturb}, we
obtain in analogy to~\eqref{Efrak} the nonlinear mapping
\[ \Delta E \::\: \H_0 \times \H_0 \times \H_0 \rightarrow \R \:,\qquad
\Delta E(\phi_\text{out}, \phi_\text{in}, \xi) := \Delta E(\p \star (\phi_\text{out}
+ \phi_\text{in} + \xi))\:. \]
After expanding similar to~\eqref{multi} in a series of multilinear mappings,
\beq \label{DelEmulti2}
\Delta E = \sum_{p,q,r=0}^\infty \Delta E_{p,q,r}
(\underbrace{\phi_\text{out},\ldots, \phi_\text{out}}_{\text{$p$ arguments}},
\underbrace{\phi_\text{in}, \ldots, \phi_\text{in}}_{\text{$q$ arguments}},
\underbrace{\xi, \ldots, \xi}_{\text{$r$ arguments}}) \:,
\eeq
we carry out the Gaussian integral with the probability distribution~\eqref{gausspos},
\begin{align*}
\Delta E(\phi_\text{out}, \phi_\text{in}) &= \sum_{p,q=0} \Delta E_{p,q}(\phi_\text{out}^p, \phi_\text{in}^q)
\qquad \text{with} \\
\Delta E_{p,q}(\phi_\text{out}^p, \phi_\text{in}^q) &:=
\sum_{r=0}^\infty \int \Delta E_{p,q,r}(\phi_\text{out}^p, \phi_\text{in}^q, \xi^r)\: d\Xi\:.
\end{align*}
Now the classical scattering operator is introduced
similar as in Section~\ref{seccscatter} by~\eqref{Efrakbil} and~\eqref{Sclass2}.

We now outline a general procedure to express the classical scattering
operator in terms of Feynman diagrams. In order to get hold on the combinatorics,
it is preferable to first derive the expansion for fixed~$\xi$, and to integrate over~$\Xi$ afterwards.
Similar to~\eqref{Efrakbil}, we rewrite the multilinear operators in~\eqref{DelEmulti2} as
linear operators on classical Fock spaces,
\begin{align*}
\Delta E \:&:\: \F_\text{class} \times \F_\text{class} \times \F_\text{class} \rightarrow \R \:, \\
&\Delta E(\underbrace{\phi_\text{out} \otimes \cdots \otimes \phi_\text{out}}_\text{$p$ factors},
\underbrace{\phi_\text{in} \otimes \cdots \otimes \phi_\text{in}}_\text{$q$ factors},
\underbrace{\xi \otimes \cdots \otimes \xi}_\text{$r$ factors})
= \Delta E_{p,q,r}(\phi_\text{out}^p, \phi_\text{in}^q, \xi^r)\:.
\end{align*}
For a fixed stochastic field~$\Xi_\text{stoch} \in \F$, we now introduce the scattering
operator~$S[\Xi_\text{stoch}]$ in analogy to~\eqref{Sclass2} by
\[ S[\Xi_\text{stoch}] \,:\, \F_\text{class} \rightarrow \F_\text{class} \quad \text{with} \quad
\Delta E(\Phi_\text{out}, \Phi_\text{in}, \Xi_\text{stoch}) = \frac{1}{2}
\la \partial_t \Phi_\text{out} \,|\, S[\Xi_\text{stoch}]^\dagger \, \Phi_\text{in} \ra \:. \]
We next consider the matrix elements of~$S[\Xi_\text{stoch}]$ for product states.
Following the procedure in Proposition~\ref{prpscatter}, we obtain
\begin{align*}
\la &\phi^1_\text{out} \otimes \cdots \otimes \phi^p_\text{out} \,|\,
S[\xi_1 \otimes \cdots \otimes \xi_r]^\dagger \, \phi^1_\text{in} \otimes \cdots \otimes \phi^q_\text{in} \ra \\
&= S_{p+q+r}(\phi^1_\text{out}, \ldots, \phi^p_\text{out},
\phi^1_\text{in}, \ldots, \phi^q_\text{in}, \xi_1, \ldots, \xi_r)\:,
\end{align*}
where~$S_{p+q+r}$ are again the classical $n$-point functions.
We thus obtain an expansion in terms of tree diagrams, but now with additional external
lines labeled by~$\xi_1, \ldots, \xi_r$. The analytic expressions for the Feynman diagrams
including the combinatorial factors are again given by~\eqref{contri2}.

\begin{figure}
\begin{picture}(0,0)%
\includegraphics{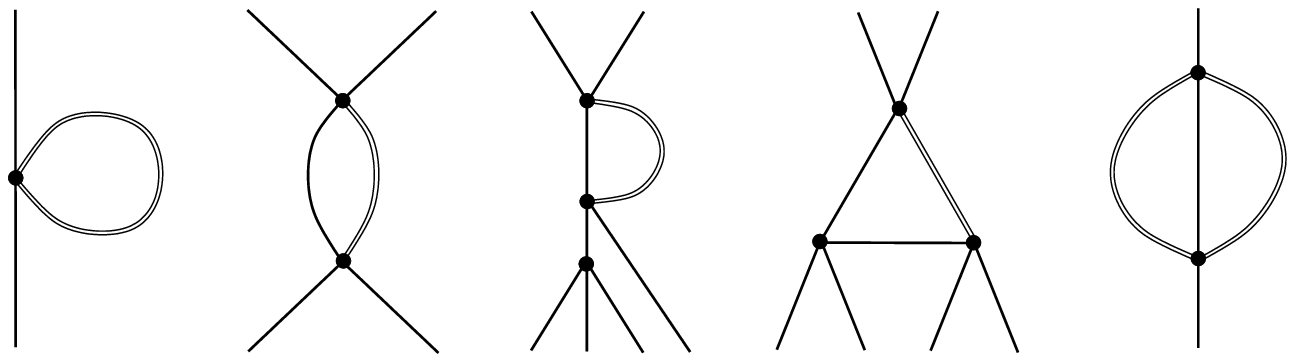}%
\end{picture}%
\setlength{\unitlength}{1533sp}%
\begingroup\makeatletter\ifx\SetFigFont\undefined%
\gdef\SetFigFont#1#2#3#4#5{%
  \reset@font\fontsize{#1}{#2pt}%
  \fontfamily{#3}\fontseries{#4}\fontshape{#5}%
  \selectfont}%
\fi\endgroup%
\begin{picture}(16185,5871)(-2542,-5147)
\put(-2527,-5006){\makebox(0,0)[lb]{\smash{{\SetFigFont{11}{13.2}{\rmdefault}{\mddefault}{\updefault}{\bf{(a)}}}}}}
\put(722,-1898){\makebox(0,0)[lb]{\smash{{\SetFigFont{11}{13.2}{\rmdefault}{\mddefault}{\updefault}$S_0$}}}}
\put(6009,-1808){\makebox(0,0)[lb]{\smash{{\SetFigFont{11}{13.2}{\rmdefault}{\mddefault}{\updefault}$P_0$}}}}
\put(4172,-1613){\makebox(0,0)[lb]{\smash{{\SetFigFont{11}{13.2}{\rmdefault}{\mddefault}{\updefault}$S_0$}}}}
\put(4262,-2783){\makebox(0,0)[lb]{\smash{{\SetFigFont{11}{13.2}{\rmdefault}{\mddefault}{\updefault}$S_0$}}}}
\put(9384,-1883){\makebox(0,0)[lb]{\smash{{\SetFigFont{11}{13.2}{\rmdefault}{\mddefault}{\updefault}$P_0$}}}}
\put(7532,-1778){\makebox(0,0)[lb]{\smash{{\SetFigFont{11}{13.2}{\rmdefault}{\mddefault}{\updefault}$S_0$}}}}
\put(8312,-2483){\makebox(0,0)[lb]{\smash{{\SetFigFont{11}{13.2}{\rmdefault}{\mddefault}{\updefault}$S_0$}}}}
\put(8959,318){\makebox(0,0)[lb]{\smash{{\SetFigFont{11}{13.2}{\rmdefault}{\mddefault}{\updefault}$x_5$}}}}
\put(7773,319){\makebox(0,0)[lb]{\smash{{\SetFigFont{11}{13.2}{\rmdefault}{\mddefault}{\updefault}$x_6$}}}}
\put(5351,342){\makebox(0,0)[lb]{\smash{{\SetFigFont{11}{13.2}{\rmdefault}{\mddefault}{\updefault}$x_5$}}}}
\put(3696,349){\makebox(0,0)[lb]{\smash{{\SetFigFont{11}{13.2}{\rmdefault}{\mddefault}{\updefault}$x_6$}}}}
\put(2719,354){\makebox(0,0)[lb]{\smash{{\SetFigFont{11}{13.2}{\rmdefault}{\mddefault}{\updefault}$x_3$}}}}
\put(240,373){\makebox(0,0)[lb]{\smash{{\SetFigFont{11}{13.2}{\rmdefault}{\mddefault}{\updefault}$x_4$}}}}
\put(-2406,367){\makebox(0,0)[lb]{\smash{{\SetFigFont{11}{13.2}{\rmdefault}{\mddefault}{\updefault}$x_2$}}}}
\put(6885,-4387){\makebox(0,0)[lb]{\smash{{\SetFigFont{11}{13.2}{\rmdefault}{\mddefault}{\updefault}$x_1$}}}}
\put(8040,-4380){\makebox(0,0)[lb]{\smash{{\SetFigFont{11}{13.2}{\rmdefault}{\mddefault}{\updefault}$x_2$}}}}
\put(8774,-4379){\makebox(0,0)[lb]{\smash{{\SetFigFont{11}{13.2}{\rmdefault}{\mddefault}{\updefault}$x_3$}}}}
\put(10011,-4387){\makebox(0,0)[lb]{\smash{{\SetFigFont{11}{13.2}{\rmdefault}{\mddefault}{\updefault}$x_4$}}}}
\put(2424,-2101){\makebox(0,0)[lb]{\smash{{\SetFigFont{11}{13.2}{\rmdefault}{\mddefault}{\updefault}$P_0$}}}}
\put(-1161,-2056){\makebox(0,0)[lb]{\smash{{\SetFigFont{11}{13.2}{\rmdefault}{\mddefault}{\updefault}$P_0$}}}}
\put(8327,-4969){\makebox(0,0)[lb]{\smash{{\SetFigFont{11}{13.2}{\rmdefault}{\mddefault}{\updefault}{\bf{(d)}}}}}}
\put(4118,-4411){\makebox(0,0)[lb]{\smash{{\SetFigFont{11}{13.2}{\rmdefault}{\mddefault}{\updefault}$x_1 \;\;\; \cdots \;\;\; x_4$}}}}
\put(-2385,-4411){\makebox(0,0)[lb]{\smash{{\SetFigFont{11}{13.2}{\rmdefault}{\mddefault}{\updefault}$x_1$}}}}
\put(403,-4386){\makebox(0,0)[lb]{\smash{{\SetFigFont{11}{13.2}{\rmdefault}{\mddefault}{\updefault}$x_1$}}}}
\put(2777,-4421){\makebox(0,0)[lb]{\smash{{\SetFigFont{11}{13.2}{\rmdefault}{\mddefault}{\updefault}$x_2$}}}}
\put(4673,-4976){\makebox(0,0)[lb]{\smash{{\SetFigFont{11}{13.2}{\rmdefault}{\mddefault}{\updefault}{\bf{(c)}}}}}}
\put(12137,-4947){\makebox(0,0)[lb]{\smash{{\SetFigFont{11}{13.2}{\rmdefault}{\mddefault}{\updefault}{\bf{(e)}}}}}}
\put(12283,-4395){\makebox(0,0)[lb]{\smash{{\SetFigFont{11}{13.2}{\rmdefault}{\mddefault}{\updefault}$x_1$}}}}
\put(12259,303){\makebox(0,0)[lb]{\smash{{\SetFigFont{11}{13.2}{\rmdefault}{\mddefault}{\updefault}$x_2$}}}}
\put(12542,-1891){\makebox(0,0)[lb]{\smash{{\SetFigFont{11}{13.2}{\rmdefault}{\mddefault}{\updefault}$S_0$}}}}
\put(13628,-1910){\makebox(0,0)[lb]{\smash{{\SetFigFont{11}{13.2}{\rmdefault}{\mddefault}{\updefault}$P_0$}}}}
\put(10700,-1880){\makebox(0,0)[lb]{\smash{{\SetFigFont{11}{13.2}{\rmdefault}{\mddefault}{\updefault}$P_0$}}}}
\put(1486,-5006){\makebox(0,0)[lb]{\smash{{\SetFigFont{11}{13.2}{\rmdefault}{\mddefault}{\updefault}{\bf{(b)}}}}}}
\end{picture}%
\caption{Examples of loop diagrams for the classical measurement process
in a stochastic background field.}
\label{figloop2}
\end{figure}

In order to carry out the integral over~$\Xi$, we first replace the factors~$\xi_1, \ldots, \xi_r$
by our stochastic field~$\xi$ to obtain an $r$-multilinear function in~$\xi$.
Then the~$\Xi$-integration gives a Gaussian integral, which can be carried out with the
Wick rules (see~\eqref{wick}). We thus obtain the sum over all pairings between
the factors~$\xi$, where every pair~$\xi(x) \,\xi(y)$ is replaced by the covariance~$\mathcal{C}(x,y)$
(see~\eqref{defcovariance}). In view of~\eqref{Cid}, every pairing generates an additional line
described by the distribution~$P_0(x,y)/\beta$ (where~$P_0$ is the fundamental solution~\eqref{P0def}).
We thus obtain loop diagrams as shown in Figure~\ref{figloop2}.

\subsection{Comparison of Classical and Quantum Loop Diagrams} \label{sec43}
Let us compare this expansion to the perturbation expansion of quantum field theory
as shown in Figure~\ref{figloop1}. First of all, the expansions are very similar in that
they both involve tree as well as loop diagrams. The only difference between the diagrams
is that the lines of the quantum diagrams are formed of the Feynman propagator, whereas the
lines in the classical diagrams are formed of either the Green's function~$S_0$ or the
fundamental solution~$P_0$, with a specific combinatorics.

Let us explain the significance of the parameter~$\beta$. We already noted in~\eqref{betadim}
that it is a dimensionless parameter. In the limit~$\beta \rightarrow \infty$, the covariance 
and thus the stochastic background field vanish. This corresponds to the classical limit where
no loop diagrams are present. As a diagram with~$l$ loops involves a factor~$\beta^{-l}$,
by decreasing~$\beta$ we can make the loop diagrams larger, giving a more ``quantum-like behavior.'' 
Thus the parameter~$\beta^{-1}$ plays a similar role as Planck's constant~$\hbar$ in
quantum theory. In order to make this connection precise, we again need to consider the
more general units~\eqref{genunit} where~$\hbar \neq 1$.
Comparing~\eqref{E0free} with~\eqref{Fockclass}, it follows that the scalar product~$\la \xi | \xi \ra$
has dimensions~$m \ell$. Thus in order for the functional~$F$ in~\eqref{QuE} to be dimensionless,
we need to choose
\[ \beta = \frac{\text{const}}{\hbar} \]
with a numerical constant. In this way, the free parameter~$\beta^{-1}$ which describes the
strength of the stochastic background field can be related to the parameter~$\hbar$
in quantum theory.

From the analytic point of view, replacing the Feynman propagator~$\triangle_F$
by~$S_0$ or~$P_0$ has no influence on the qualitative behavior of
the loops. In particular, the classical loop diagrams
have the same divergent behavior as the corresponding quantum diagrams.
The following lemma even gives a quantitative connection between
the classical and the quantum loops.
\begin{Lemma} \label{lemmarep}
Assume that the momenta~$k_1, \ldots, k_r$ with~$r \geq 1$ are all different. Then
for almost all momenta~$\vec{q}$,
\[ \int_{-\infty}^\infty dq^0 \,\triangle_F(k_1+q) \cdots \triangle_F(k_r+ q)
= -i \pi \int_{-\infty}^\infty dq^0 \:\sum_{a=1}^r P_0(k_a+q)\, \prod_{b \neq a} S_0(k_b+q) \:. \]
\end{Lemma}
\Proof We set~$\omega=q^0$, $\omega_a=k_a^0$ and~$p_a=|\vec{k_a}+\vec{q}|>0$. Then,
according to~\eqref{Feynprop},
\begin{align}
\triangle_F(k_a+q) &= \lim_{\varepsilon \searrow 0} \frac{1}{(\omega+\omega_a)^2 - p_a^2 + i \varepsilon} 
\label{D1} \\
&= \lim_{\varepsilon \searrow 0} \frac{1}{2p_a} \bigg( \frac{1}{(\omega+\omega_a) - p_a + i \varepsilon}
- \frac{1}{(\omega+\omega_a) + p_a - i \varepsilon} \bigg) \:. \label{D2}
\end{align}
We take the product of~$r$ such factors for fixed~$\varepsilon>0$.
Since the momenta~$k_1, \ldots, k_r$ are pairwise different, the resulting function has simple
poles for almost all~$\vec{q}$. In the case~$r \geq 2$, we can close the contour either in the
upper or in the lower half plane.
Computing the mean value of these two contour integrals and taking the limit~$\varepsilon \searrow 0$,
we obtain
\begin{align*}
\int_{-\infty}^\infty & dq^0 \,\triangle_F(k_1+q) \cdots \triangle_F(k_r+ q)
= -i \pi \sum_{a=1}^r \frac{1}{2 p_a} \prod_{b \neq a} S_0(k_b+q) \big|_{\omega =-\omega_a \pm p_a} \\
&=  -i \pi \int_{-\infty}^\infty d\omega
\sum_{a=1}^r \frac{1}{2 p_a} \Big( \delta(\omega + \omega_a + p_a) + \delta(\omega+\omega_a - p_a)
\Big) \,\prod_{b \neq a} S_0(k_b+q) \\
&=  -i \pi \int_{-\infty}^\infty d\omega
\sum_{a=1}^r \delta \big( (k_a+q)^2 \big) \,\prod_{b \neq a} S_0(k_b+q) \:.
\end{align*}
Using~\eqref{P0def} gives the result.
In the case~$r=1$, the two summands in~\eqref{D2} do not decay fast enough for closing
the contours. However, one can compute the contour integrals using~\eqref{D1},
to again obtain the result.
\QED
This lemma shows that in a closed loop, a product of $r$ Feynman propagators may be replaced
by the product of one factor~$P_0$ and $r-1$ factors~$S_0$, if we take the sum over all such
combinations. The condition that the momenta~$k_1, \ldots, k_r$ should all be different
can be justified as follows: When computing scattering amplitudes, the momenta~$k_1, \ldots, k_r$
are expressed in terms of the incoming and outgoing momenta. Then the momenta~$k_a$
are are all different except at the poles of the scattering amplitudes.
Such poles correspond to resonances; they need to be described non-perturbatively
with a resummation technique.
If the loop under consideration is contained inside another loop, one also needs to integrate
over one or several of the momenta~$k_a$. In this case, the poles of the Feynman diagrams
if some of the momenta~$k_1, \ldots, k_r$ coincide are of no relevance, as only the
divergences for large momenta are of interest.

Thus, in order to decide whether the classical and the
quantum loop diagrams agree quantitatively, it remains to study the combinatorics.
Let us begin with the tadpole diagram (see Figure~\ref{figloop1}~{\bf{(b)}}).
It arises when integrating the simple tree diagram (see the left of Figure~\ref{figtree2})
over~$\Xi$, if two of the incoming fields are the stochastic field~$\xi$.
As there are three possibilities for the choice of the two $\xi$-legs, 
and taking into account the factor~$1/6$ in~\eqref{perturb}, we get a total combinatorial
factor of~$1/2$. Hence Lemma~\ref{lemmarep} allows us to replace
the factor~$P_0$ in the classical tadpole by
\[ P_0 \rightarrow -\frac{1}{2 \pi i}\: \triangle_F\:. \]
As the factor~$(-i)$ is already taken care of in~\eqref{contri}, we get complete agreement
between the quantum and the classical tadpole if we choose (see~\eqref{defcovariance} and~\eqref{Cid})
\beq \label{beta1}
\beta = \frac{1}{2 \pi}\:.
\eeq

We next consider the simple loop diagram in Figure~\ref{figloop1}~{\bf{(d)}}.
In the classical setting, this diagram arises by performing the stochastic integral
of a second order tree diagram
with two $\xi$-legs (see Figure~\ref{figloop2}~{\bf{(b)}}).
As the tree diagram has a combinatorial factor one (see Proposition~\ref{prpscatter}),
the inner lines are formed by the product~$\lambda^2 S_0 P_0$.
Due to the symmetry in the momenta of the two inner lines, Lemma~\ref{lemmarep}
allows us to replace the inner lines by~$\triangle_F \triangle_F/(-2 \pi i)$.
Hence we again get complete agreement between the quantum and the classical diagram
if we choose~$\beta$ according to~\eqref{beta1}.
This argument applies just as well to simple loops inside larger diagrams,
as is exemplified in Figure~\ref{figloop2}~{\bf{(c)}}.

The situation changes with longer loops, as we now discuss in the example
of the triangle loop in Figure~\ref{figloop2}~{\bf{(d)}}. The shown diagram is obtained in a unique way
as the stochastic integral of a third order tree diagram with two $\xi$-legs.
The combinatorial factor is again one. Similarly, one gets the two loop diagrams where
the line~$P_0$ is one of the other sides of the triangle. Thus we can again apply
Lemma~\ref{lemmarep}. In order to obtain complete agreement to the corresponding quantum
diagram we need to choose
\beq \label{beta2}
\beta = \frac{1}{\pi}\:.
\eeq
We point out that this differs from~\eqref{beta1} by a factor of two.
This shows that the expansion in terms of classical loop diagrams is not mathematically
equivalent to quantum field theory. One difference is that longer loops have a weight
which is double that of a tadpole or a simple loop.

For higher loops (see Figure~\ref{figloop2}~{\bf{(e)}} for the simplest example),
one can apply inductively Lemma~\ref{lemmarep}
(or a generalization where on the left factors~$\triangle_F$ are replaced by~$S_0$)
to get a connection between the classical and the corresponding quantum diagrams.
However, the combinatorial details of this analysis go beyond the scope of the present paper.

\subsection{The Zero-Point Energy of the Stochastic Field} \label{seczero}
A quantum field has an infinite zero-point energy, which is detectable via the Casimir effect.
The analogous quantity in our classical setting is the mean energy of the classical stochastic field~$\xi$,
given formally by
\beq \label{zpe}
\int (\xi, \xi) \:\D \Xi \:.
\eeq
We point out that following our discussion in Section~\eqref{secmeasure}, this
``zero point energy'' is not directly accessible to an observer, as
only energy differences can be measured.
Nevertheless, it is conceivable that this energy could be measured in more realistic
physical models which include fermions and thus make it possible to set up the Casimir experiment.
With this in mind, it seems interesting to compute~\eqref{zpe} and to compare the result
with the zero-point energy of a quantum field.

The mean energy of the stochastic field~\eqref{zpe} is infinite for two reasons: First, the
mean energy density is infinite in view of~\eqref{defcovariance} and~\eqref{Cid},
taking into account that the distribution~$P_0(x,y)$ is singular at~$x=y$.
Second, the integral of the energy density diverges because the spatial volume is infinite.
In order to avoid the second problem, we consider the system as usual in
a three-di\-men\-sio\-nal box of length~$L$ with periodic boundary conditions.
Then the Fourier integral~\eqref{P0def} is to be replaced by the Fourier series
\[ P_0(x,y) = \frac{1}{2 \pi L^3} \sum_{\vec{k} \in (2 \pi \Z/L)^3} 
\frac{1}{2 |\vec{k}|} \left( e^{-i \omega t} + e^{i \omega t} \right) \:e^{i \vec{k} (\vec{x}-\vec{y})} \:, \]
where we set~$\omega = |\vec{k}|$ and~$t = x^0 - y^0$.
Using~\eqref{Cid} and~\eqref{E0free}, we obtain
\begin{align*}
\int (\xi, \xi) \:\D \Xi &= \frac{1}{\beta}\:
\frac{1}{2 \pi L^3} \sum_{\vec{k} \in (2 \pi \Z/L)^3} 
\frac{1}{2 |\vec{k}|} \sum_{s,s' = \pm 1}
\frac{1}{2} \left( - s s' \,\omega^2 + |\vec{k}|^2 \right) L^3 \\
&= \frac{1}{4 \pi \beta} \sum_{\vec{k} \in (2 \pi \Z/L)^3}
\sum_{s = \pm 1} \omega = \frac{1}{2 \pi \beta} \sum_{\vec{k} \in (2 \pi \Z/L)^3} \omega\:.
\end{align*}
Here the infinite sum clearly diverges. But to every
momentum mode we can associate the mean energy
\[ E = \frac{\omega}{2 \pi \beta}\:. \]
Choosing~$\beta$ according to~\eqref{beta2} gives perfect agreement with the
zero point energy~$\hbar \omega/2$ of a radiation mode of a quantum field
as described by a quantum mechanical harmonic oscillator.

\section{Outlook} \label{sec5}
The previous analysis shows that by considering the classical $\phi^4$-theory in
a stochastic background field, one gets a perturbative description which
involves all the tree and loop diagrams of the corresponding quantum field theory.
The main result of the paper is to demonstrate that loop diagrams appear
in a purely classical context. By adapting our methods, this result immediately carries
over to any other bosonic field theory in any space-time dimension.

We also observed that for $\phi^4$-theory, the combinatorics of the tadpole and of the
simple loops differs from that of the longer loops (see Section~\ref{sec43}).
This result is quite sensitive on the
theory under consideration. It might well be that for other theories, these combinatorial
differences between the classical and the quantum loop diagrams disappear.
In particular, it was crucial for our combinatorial considerations that there are distinct lines
which begin at the same vertex~$i$ and both end at the same vertex~$j$.
As such lines do not exist in QED, is conceivable that for QED there might be full agreement
between classical and the quantum loop diagrams.

In any case, working out the combinatorial details for a realistic model to higher order seems an
important project for the future. Here we only make three remarks:
\begin{itemize}
\item[(a)] We note that for complex fields (like a complex Klein-Gordon field), the charge is an additional
conserved quantity. This makes it possible to work in the classical measurement process
instead of the energy with the charge. An advantage of this procedure is that the charge is a Lorentz
scalar, so that the time derivative in~\eqref{Sclass2} could be omitted.
The reason why we preferred to work with a real field is that the components of the
electromagnetic field are also real-valued.
\item[(b)] We remark that in order to extend our methods, one could replace the
Gaussian measure~\eqref{gausspos} by a more general probability distribution on
the classical solutions. This would change the detailed form of the higher order loop diagrams.
The non-Gaussian probability distribution would have to be justified as arising from
the nonlinear classical dynamics prior to the scattering process.
\item[(c)] We point out that all our methods only work for {\em{bosonic}} fields. The reason is that
the Pauli exclusion principle 
and the transformation of half-integer spin fields under rotations
seem to make it impossible to
interpret the Dirac equation as a classical field equation.
Thus in order to extend our methods to models involving fermions, one possibility
is to couple a classical bosonic field to a second-quantized fermion field.
An alternative is to describe the fermions in the framework of the fermionic projector~\cite{srev}.
\end{itemize}

We finally outline how our constructions could be helpful for the rigorous construction of interacting
quantum fields. To this end, let us assume that there is a physical theory for which
there is complete agreement between the classical and the quantum Feynman diagrams.
Then one could prove the existence of the quantum theory by rigorously constructing the
classical theory in the presence of a stochastic background field.
More precisely, one could first introduce an ultraviolet regularization by replacing the
covariance of the Gaussian measure~\eqref{Cid} by a regularized covariance~$P^\varepsilon/\beta$
(for example with a cutoff in momentum space; clearly, the regularization would violate
Lorentz invariance). Then the initial values $\phi_0|_{t=-T}$ (with~$\phi_0$ according
to~\eqref{phi0tot}) would be in a suitable Sobolev space.
If one could prove global existence of solutions uniformly in~$\xi$, integrating the solutions
over~$\Xi$ would give all interacting $n$-point functions. If one also succeeded in
controlling the behavior of the classical solutions in the limit when the regularization is removed
(this would probably require some kind of ``non-perturbative renormalization method''), one would have proved
non-perturbatively that the quantum field theory in the Lorentzian setting is well-defined and finite.
In this way, it seems possible at least in principle to make the powerful methods of nonlinear
partial differential equations applicable to one of the most outstanding problems of quantum field
theory.

\Thanks{{\em{Acknowledgments:}} We would like to thank Eberhard Zeidler 
for inspiring and encouraging discussions. We are grateful to Andreas Grotz, Olaf M\"uller
and the referee for helpful comments on the manuscript.} \\[.5cm]

\bibliographystyle{amsplain}
%\bibliography{../felix}

\def\dbar{\leavevmode\hbox to 0pt{\hskip.2ex \accent"16\hss}d}
\providecommand{\bysame}{\leavevmode\hbox to3em{\hrulefill}\thinspace}
\providecommand{\MR}{\relax\ifhmode\unskip\space\fi MR }
% \MRhref is called by the amsart/book/proc definition of \MR.
\providecommand{\MRhref}[2]{%
  \href{http://www.ams.org/mathscinet-getitem?mr=#1}{#2}
}
\providecommand{\href}[2]{#2}

\end{document}